\documentclass[aps,pre,twocolumn]{revtex4}
\usepackage[dvips]{graphics}
\usepackage{graphicx}
\usepackage{amsfonts}
\usepackage{amssymb}
\usepackage{amsmath}
\usepackage{natbib}
\usepackage{color}

\begin{document}

\title{Manipulation of Vortices by Localized Impurities in Bose-Einstein Condensates}

\author{M.C.~Davis$^1$, R~Carretero-Gonz\'{a}lez$^1$, Z.~Shi$^2$, K.J.H.~Law$^2$,
P.G.~Kevrekidis$^2$, and B.P.~Anderson$^3$}
\affiliation{$^1$Nonlinear Dynamical Systems Group\footnote{URL: {\tt http://nlds.sdsu.edu/}},
Department of Mathematics and Statistics,
and Computational Science Research Center\footnote{URL: {\tt http://www.csrc.sdsu.edu/}},
San Diego State University, San Diego CA, 92182-7720, USA.\\
$^2$Department of Mathematics and Statistics,
University of Massachusetts, Amherst MA 01003,USA \\
$^3$College of Optical Sciences and Department of Physics, University of Arizona,
Tucson AZ, 85721, USA.}

\begin{abstract}
We consider the manipulation of Bose-Einstein condensate
vortices by optical potentials generated by focused laser beams.
It is shown that for
appropriate choices of the laser strength and width it is
possible to successfully transport vortices to various
positions inside the trap confining the condensate atoms.
Furthermore,
the full bifurcation structure of possible stationary single-charge
vortex solutions in a harmonic potential with this type of impurity
is elucidated.
The case when a moving vortex is
captured by a stationary laser beam is also studied,
as well as the possibility of dragging the vortex
by means of periodic optical lattices.

\end{abstract}

\date{\today
}
\maketitle

\section{Introduction}

Interactions between localized impurities, or pinning centers, and flux lines in type-II superconductors have long been of interest in condensed matter physics \cite{Anderson1962aCampbell1972aDaldini1974aCivale1991a}, with much recent work focusing on the pinning effects of arrays of impurities \cite{Baert1995aReichhardt2001aGrigorenko2003a}.  Similar studies of the interactions between a vortex array in a rotating Bose-Einstein condensate (BEC) and a co-rotating optical lattice  \cite{Reijnders2004aPu2005aReijnders2005a} have further contributed to the interest in the physics of manipulating one array of topological structures with a second array of pinning sites.  Depending on the configuration, depth, and rotation rate of the optical lattice, structural changes to the vortex array may be induced, and have now been experimentally observed \cite{Tung2006a}. Furthermore, combining an optical lattice with a rotating BEC may enable new investigations of other interesting phenomena, such as for example, alterations to the superfluid to Mott-insulator transition \cite{Bhat2006aGoldbaum2008a}, production of vortex liquids with broken translational symmetry \cite{Dahl2008a}, and the existence of stable vortex molecules and multi-quantum vortices \cite{Geurts2008a}.  Yet despite these significant advances, the interactions between a \emph{single} vortex and a single pinning site within a BEC, and the associated vortex dynamics, are not fully understood and many problems remain unexplored.  A more complete understanding of such basic interactions may be important for the further development of many ideas and experiments regarding vortex pinning and manipulation, even for the case of vortex arrays.  Here we undertake a theoretical and numerical study that examines the possibility of vortex capture and pinning at a localized impurity within the BEC, and the possibility of vortex manipulation and dragging by a moving impurity.

Manipulation of coherent nonlinear matter-wave structures \cite{BECBOOK,NonlinearityReview} in trapped BECs has indeed received some examination \cite{Manipulation-SPIE}. For example, in the case of negative scattering length (attractive) BECs in
a quasi-one-dimensional (1D) scenario, numerical analysis shows that it is possible to
pin bright solitons away from the center of harmonic trap.
More importantly, pinned bright solitons may be adiabatically dragged and repositioned
within the trap by slowly moving
an external impurity generated by a focused laser beam \cite{BS-Imp}.
Alternatively, bright solitons might be pinned and dragged
by the effective local minima generated by adiabatically
moving optical lattices and superlattices \cite{BS-OL1,BS-OSL2}.
The case of repulsive interactions has also drawn considerable
attention. In the 1D setting, the effect of localized impurities
on dark solitons was described in Ref.~\cite{DS-Imp}, by using direct
perturbation theory \cite{vvkve}, and later in Ref.~\cite{fr1},
by the adiabatic perturbation theory for dark solitons \cite{KY}.
Also, the effects and possible manipulation of dark solitons
by optical lattices have been studied in
Refs.~\cite{DS-OL1,DS-OL2,DS-OL3}.

In the present work, we limit our study of vortex-impurity interactions and
vortex manipulation to the case of a positive scattering length (repulsive) pancake-shaped BEC that is harmonically trapped. We envision a single localized impurity created by the addition of a focused laser beam \cite{BECBOOK},
which may in principle be tuned either above or below the atomic resonance, thereby creating a repulsive or attractive potential with blue or red detunings, respectively.  We concentrate on the dynamics of a blue-detuned beam
interacting with a single vortex.

Our manuscript
is organized as follows. In the next section we
describe the physical setup and its mathematical model.
In Sec.~\ref{SEC:static} we study the static scenario
of vortex pinning by the localized laser beam by
describing in detail the full bifurcation
structure of stationary vortex solutions and their
stability as a function of the laser properties and
the pinning position inside a harmonic trap.
In Sec.~\ref{SEC:dragging} we study vortex
dragging by an adiabatically
moving impurity. We briefly describe our observations also for the
case of single vortex manipulation using an optical lattice, and touch upon the
possibility of capturing a precessing
vortex by a fixed impurity.
Finally, in Sec.~\ref{SEC:conclu} we summarize our results
and discuss some possible generalizations and open
problems.

\section{Setup\label{SEC:setup}}

In the context of BECs at nano-Kelvin temperatures,
mean-field theory can be used to accurately approximate the behavior of
matter-waves \cite{BECBOOK}. The resulting mathematical model is a
particular form of the nonlinear Schr\"odinger equation (NLS) known as the
Gross-Pitaevskii equation (GPE) \cite{Gross:61,Pitaevskii:61}.
The GPE in its full dimensional form is as follows:
\begin{equation}
i\hbar \psi_{t}=-\frac{\hbar^{2}}{2m} \nabla^{2} \psi
+ g |\psi|^{2}\psi + V(x,y,z,t)\,\psi, \label{dimGPE}
\end{equation}
where $\psi(x,y,z,t)$
is the wavefunction describing the condensate,
$m$ is the mass of the condensed atoms, $g={4\pi \hbar^{2}a}/{m}$
and $a$ is their $s$-wave
scattering length. The time-dependent external potential
$V(x,y,z,t)$ acting on the condensate is taken to be a combination of
a static harmonic trap (HT) holding the condensed atoms, and a localized
impurity (Imp) provided by a narrowly focused laser beam:
\begin{equation}
V(x,y,z,t) = V_{\rm HT}(x,y,z) + V_{\rm Imp}(x,y,z,t). \label{V3D}
\end{equation}
Herein we consider a harmonic trap potential
\begin{equation}
V_{\rm HT}(x,y,z) = \frac{m}{2} \omega_r^2\left(x^{2}+y^{2}\right)
+ \frac{m}{2} \omega_z^2 z^{2}, \label{HT0}
\end{equation}
with trapping frequencies $\omega_r$ and $\omega_z$ in the radial
and $z$ directions respectively.
In general, $V_{\rm Imp}$ can be a negative or positive quantity,
corresponding to an impurity that is an attractive or
repulsive potential for the trapped atoms.

In the present study we further limit our attention to
quasi-two-dimensional condensates,
the so-called pancake-shaped condensates, by considering
that
$\omega_z \gg \omega_r$
and that the tight ($z$) direction
condensate profile is described by the harmonic trap ground state
in that direction \cite{BECBOOK,NonlinearityReview}.
We also consider only cases where $V_{\rm Imp}$ is only a function of $x$ and $y$,
and possibly $t$, and hereafter remove the $z$ dependence from our notation.
Under this assumption it is possible
to reduce the three-dimensional GPE (\ref{dimGPE}) to an
effective two-dimensional equation that has the same form as
its three-dimensional counterpart but with $g$
replaced by $g_{\rm 2D}=g/\sqrt{2\pi} a_z$, where $a_z=\sqrt{\hbar/(m\omega_z)}$
is the transverse harmonic oscillator length \cite{BECBOOK,NonlinearityReview}.
%
%
%\bpa{In the following, I'm concerned about the density units: $\sqrt{\hbar\omega_z/g_{\rm 2D}}$ has dimension of inverse length, but the 2D density that we're working with has units of inverse length squared once the renormalization is done.  Also, isn't it the 2D wavefunction that should be used in the above sentence rather than $|\psi|^2$, since $\psi$ is the full 3D wavefunction?  I think the next sentence could be modified by removing the square-root in $\sqrt{\hbar\omega_z/g_{\rm 2D}}$, and by saying ``the two-dimensional density'' and just leaving out the $|\psi|^2$ from the sentence.  Please check to see if what I'm suggesting makes sense.}
%
%\bpa{Also: should we also explicitly state that energies are in units of $\hbar\omega_z$?}
%
%{\bf kjhl:  BPA is exactly right, and with good suggestions.  I implement them below.}

Furthermore, by measuring, respectively, two-dimensional density,
%$|\psi|^2$,
length, time, and energy in units of
%$\sqrt{2}/(4\pi a a_z)$, $a_z$, $\omega_z^{-1}$, and $\hbar \omega_z$,
%rcg: I re-did renormalization:
$\hbar\omega_z/g_{\rm 2D}$, $a_z$, $\omega_z^{-1}$, and $\hbar \omega_z$,
one obtains the standard
form for the adimensionalized GPE in two dimensions:
\begin{equation}
iu_{t}=-\frac{1}{2}(u_{xx} + u_{yy}) + |u|^{2}u + V(x,y,t)u, \label{GPE}
\end{equation}
where the harmonic potential now reads
\begin{equation}
V_{\rm HT}(x,y) = \frac{\Omega^2}{2}\left(x^{2}+y^{2}\right), \label{HT}
\end{equation}
and $\Omega \equiv \omega_r/\omega_z$
is the adimensionalized harmonic trap strength.
We use throughout this
work a typical value for the harmonic trap strength of $\Omega=0.065$
unless stated otherwise.
Other harmonic trap strengths gave %similar qualitative
qualitatively similar results.
In addition to the harmonic trap we %will
impose a localized potential
stemming from an external localized laser beam centered
about $(x_a(t),y_a(t))$ that in adimensional
form reads
\begin{equation}
V_{\rm Imp}(x,y,t) = V^{(0)}_{\rm Imp}\ \exp\left(-\frac{
  [x-x_a(t)]^{2}+[y-y_a(t)]^{2}}{\varepsilon^{2}}\right). \label{Imp}
\end{equation}
In this equation,  $V^{(0)}_{\rm Imp}$
is proportional to the peak laser
intensity divided by the detuning of the laser from the atomic resonance,
and $\varepsilon = w_0/\sqrt{2}$ where $2w_0$ is the adimensional Gaussian beam width.
A positive (negative)
$V^{(0)}_{\rm Imp}$ corresponds to the intensity of a blue-(red-)detuned,
repulsive (attractive) potential.

%\section{Steady State Solutions}

Steady-state solutions of the GPE are obtained by separating spatial and
temporal dependencies as $u(x,y,t)=\Psi(x,y)\ e^{-i\mu t}$,
where $\Psi$ is the steady-state, time-independent amplitude of the wavefunction
and $\mu$ is its chemical potential (taken here as $\mu=1$ in
adimensional units).
Under the conditions that the location of the impurity is time-independent such that $x_a$ and $y_a$ are constant, this leads to the steady-state equation
\begin{equation}
\mu\Psi = -\frac{1}{2}\left(\Psi_{xx} + \Psi_{yy}\right)+|\Psi|^{2}\Psi + V(x,y)\Psi.
\label{steadyst}
\end{equation}

The initial condition used in this study was one that
closely approximates a vortex solution of unit charge $s=\pm 1$
centered at $(x_{0},y_{0})$:
\begin{equation}
\begin{array}{rcl}
\Psi(x,y) &=& \Psi_{\rm TF}(x,y)\, \tanh[(x-x_{0})^{2}+(y-y_{0})^{2}] \\[2.0ex]
&& \times\exp\left[is\ \tan^{-1}\left\{(y-y_{0})/(x-x_{0})\right\}\right],
\end{array}
\end{equation}
where $\Psi_{\rm TF}(x,y) = \sqrt{\max(\mu-V(x,y),0)}$
represents the shape of the Thomas-Fermi (TF) cloud formed in the presence of
the
relevant external potentials \cite{BECBOOK}.
%, and the inverse tangent spans a full four quadrants so that all angles between 0 and $2\pi$ may be obtained.
%
%
Subsequently, this approximate initial condition was
allowed to converge to the numerically ``exact'' solutions
by means of fixed point iterations.

\section{The Static Picture: Vortex Pinning and the bifurcations beneath}
\label{SEC:static}

\subsection{Vortex Pinning by the Impurity}

It is well-known that a vortex interacting with a harmonic trap undergoes
a precession based upon the healing length of the vortex and the
parameters which define the trap \cite{precession1,precession2,precession3,precession4,precession5,precession6,precession7,precession8}.
Since we are introducing a localized
impurity into the trap it is
worthwhile to first observe the behavior of a
vortex interacting with only the localized impurity,
in the absence of the harmonic potential.
We note that in this case, the parameters $x_a$ and $y_a$ may be
neglected and the parameters $x_0$ and $y_0$ can be interpreted
as the coordinates of the vortex relative to the impurity.
%

%%%%%%%%%%%%%%%%%%%%%%%%%%%%%%%%%%%%%%%%%%%%%%%%%%%%%%%%%%%%%%%%%%%%%%%%
\begin{figure}[ht]
%\newline
  \centerline{
    \includegraphics[width=6.5cm,angle=-90,clip]{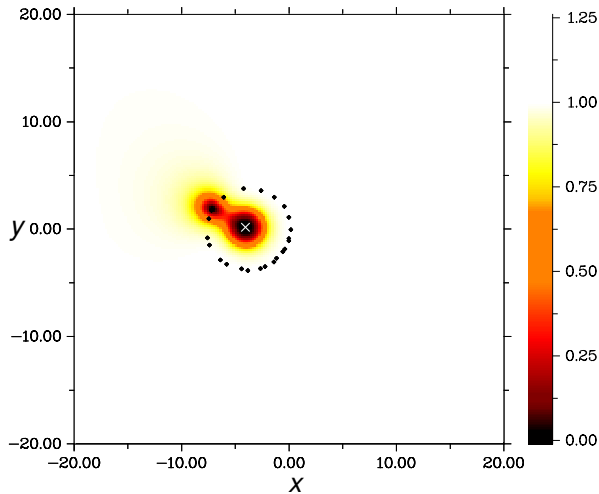}
  }
  \caption{(Color online) Density plot showing a snapshot of the interaction
    of the vortex with a localized impurity in the absence of
    the harmonic trap (i.e., $\Omega = 0$).
    The presence of the impurity (wider field depression) induces a
    clockwise rotation of the vortex (narrower field depression)
    along a path depicted by the dark dots.
    The parameters are as follows:
    $(\mu,\Omega,s,V_{\rm Imp}^{(0)},\varepsilon)=(1,0,1,5,1)$.     %,18)$
    The colorbar shows the condensate density in adimensional units.
  }
  \label{man_18}
\end{figure}
%%%%%%%%%%%%%%%%%%%%%%%%%%%%%%%%%%%%%%%%%%%%%%%%%%%%%%%%%%%%%%%%%%%%%%%%

%%%%%%%%%%%%%%%%%%%%%%%%%%%%%%%%%%%%%%%%%%%%%%%%%%%%%%%%%%%%%%%%%%%%%%%%
\begin{figure}[ht]
  %\newline
  \centerline{
    \includegraphics[width=9.cm,height=7cm,clip]{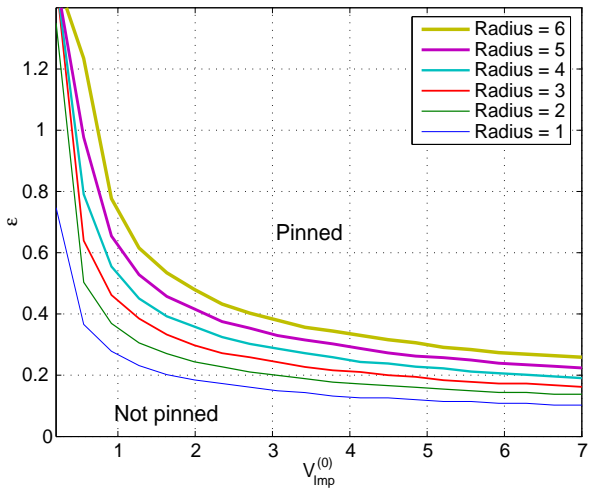}
  }
  \caption{(Color online) Phase diagram depicting the vortex pinning by the
    localized impurity of strength $V_{\rm imp}^{(0)}$ and
    width $\varepsilon$. Each curve represents a different pinning
    location at the indicated radii (i.e., distance from the
    center of the harmonic trap).
    Parameters are as follows: $(\mu,\Omega,s)\ =\ (1,0.065,1)$.
  }
  \label{pin_exist}
\end{figure}
%%%%%%%%%%%%%%%%%%%%%%%%%%%%%%%%%%%%%%%%%%%%%%%%%%%%%%%%%%%%%%%%%%%%%%%%

By symmetry, a vortex placed at the center of an impurity
(i.e., $x_0 = y_0 = 0$) will result
in a steady state without precessing. However, a vortex placed off center with
respect to the impurity will precess at constant speed
around the impurity due to the
gradient in the background field induced by the impurity \cite{Kivshar98}.
In order to study this behavior in a simple physically
meaningful
setting
we start with a
positive-charge ($s=1$)
vortex without the impurity and then the
impurity is adiabatically
switched on at a prescribed distance away from the
center of the vortex. We find that for $V_{\rm Imp}^{(0)}>0$
the vortex then begins to precess around the impurity in a
clockwise direction.  Reversing the sign of $V_{\rm Imp}^{(0)}$
in order to create an attractive impurity
induces
a counter-clockwise precession with respect to the impurity.
%
%We will see later that this is the same direction of precession that is
%induced by the harmonic trap.  Thus for repulsive impurities of the sort
%that we concentrate on in the following, there is an effective competition
%between the precession induced by the impurity and that induced by the harmonic potential.
%
An example of the vortex precession induced by the impurity is
shown in Fig.~\ref{man_18}.
It is crucial to note that
if the impurity is turned on
``close enough'' to the steady-state vortex such
that the impurity is within the vortex funnel then the vortex would begin its
usual rotation but would be drawn into the center of the impurity, effectively
pinning the vortex.
This effective attraction is related to the emission
of sound by the vortex when it is inside the
funnel of the impurity as described in Ref.~\cite{Parker:04}.

Throughout this work we follow the center of the vortices by
detecting the extrema of the superfluid vorticity $\mbox{\boldmath$\omega$}$
defined as $\mbox{\boldmath$\omega$}=\nabla\times{\mathbf v}_s$ where
the superfluid velocity
in dimensional units
is \cite{Jackson:98,BECBOOK}
\begin{equation}
{\mathbf v}_s=-\frac{i \hbar}{2m} \frac{\psi^*\nabla \psi - \psi\nabla \psi^*}{|\psi|^2},
\label{fluid_vel}
\end{equation}
where $(\cdot)^*$ stands for complex conjugation.

We now consider the net effect of the pinning induced by the impurity
and the precession induced by the harmonic trap. Since one of our
main goals is to find
conditions needed for the manipulation of
vortices using the repulsive impurity, a \emph{minimum} requirement would
be that the impurity's pinning strength is sufficient to overcome the
precession inside the trap
and thus pin the vortex very close to the location of the impurity
(i.e., $(x_0,y_0) \approx (x_a,y_a)$).
%
%Therefore, we seek to find the
%minimum conditions that an off-center vortex, at a particular radius measured
%from the center of the harmonic trap, could be pinned by a
%localized impurity
%at position $(x_a, y_a)$.
Therefore, we seek to find the minimum conditions that an off-center vortex, at a particular radius measured from the center of the harmonic trap, could be pinned by a localized impurity at that same location.  For certain combinations of beam parameters (strong beam intensity, or large beam widths), the vortex will remain localized near this point.  For other parameters (weak intensity, small beam widths), the beam cannot overcome the vortex precession induced by the harmonic trap and the vortex would not remain localized near the beam position.
This would give us a lower bound for the possible beam intensities and widths
for which a vortex might be dragged to the corresponding position
within the BEC.
The existence of such pinned states was  identified by searching in the
impurity parameter space (strength $V_{\rm imp}^{(0)}$ and
width $\varepsilon$) for several off-center radii, i.e., distances
measured from the center of the trap.
The results are shown in the
phase diagrams of Fig.~\ref{pin_exist} where each curve corresponding
to a different radius (decreasing from top to bottom) depicts the
boundary in $(V_{\rm imp}^{(0)},\varepsilon)$ parameter space for which
pinning is possible.  In other words, for the points in parameter space below a
given curve, one gets primarily vortex precession dynamics induced
by the harmonic trap, whereas above these curves (i.e., for
strong or wide enough impurities), the vortex
is trapped by the impurity and stays very close to it.

\subsection{Steady-state bifurcation structure\label{SEC:bifurcations}}

%%%%%%%%%%%%%%%%%%%%%%%%%%%%%%%%%%%%%%%%%%%%%%%%%%%%%%%%%%%%%%%%%%%%%%%%%

%\begin{figure}[tbp]
%  %\newline
%  \centerline{
%    \includegraphics[width=11.cm,angle=0]{./bif_scheme.eps}
%  }
%  \caption{(Color online) A cartoon of a slice of the potential
%with $y=0$ is shown for fixed $V^{(0)}_{\rm{Imp}}$ as the radius
%of the impurity offset, $x_a$ varies between zero (left panel)
%and $x_{a,\rm{cr}}$ (right panel) -
%the critical radius for pinning (shown by a red line).  The
%critical points of the potential  are indicated by red circles and
%represent the location of the vortex in the corresponding
%stationary vortex state for given $x_a$.
%The additional symmetry for $x_a=0$ implies
%that at the two minima of the potential the solutions are
%rotated versions of the same solution
%%due to the additional symmetry
%(as are any of the equivalent solutions for the one-parameter family
%$\theta \in (0,2\pi]$, in
%radial coordinates, as opposed to only $\theta \in \{0,1\}$ as in the
%diagram). There are three solutions for any given radius between
%(and not including) zero and $x_{a,\rm{cr}}$. At either end, there exists
%a transcritical and a saddle-node bifurcation, respectively.
%}
%  \label{bif_scheme}
%\end{figure}

%%%%%%%%%%%%%%%%%%%%%%%%%%%%%%%%%%%%%%%%%%%%%%%%%%%%%%%%%%%%%%%%%%%%%%%%%%%%%

%%%%%%%%%%%%%%%%%%%%%%%%%%%%%%%%%%%%%%%%%%%%%%%%%%%%%%%%%%%%%%%%%%%%%%%%%

\begin{figure}[tbp]
  \centerline{
   \includegraphics[width=7.0cm,angle=0]{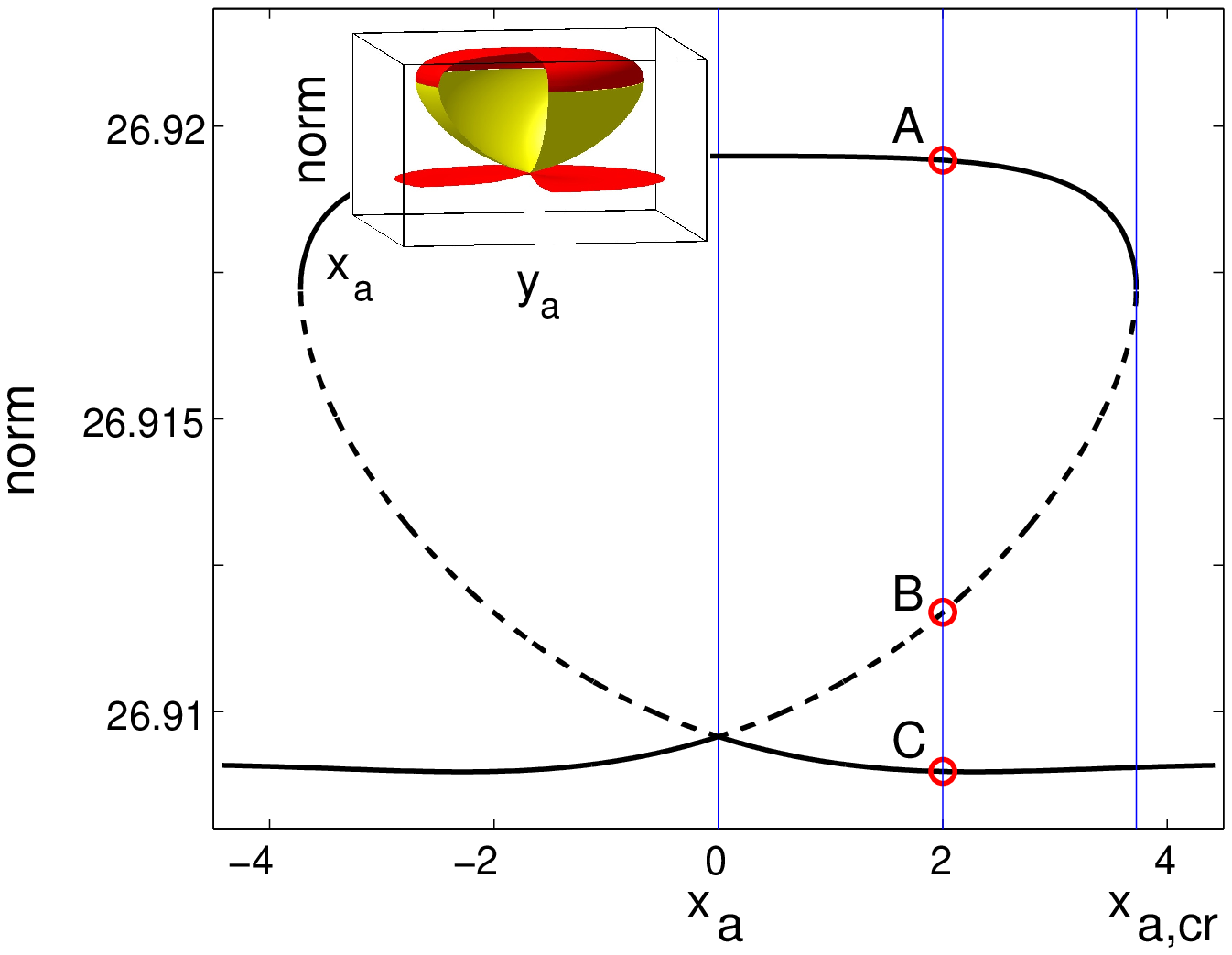}
}
   \includegraphics[width=4.0cm,angle=0]{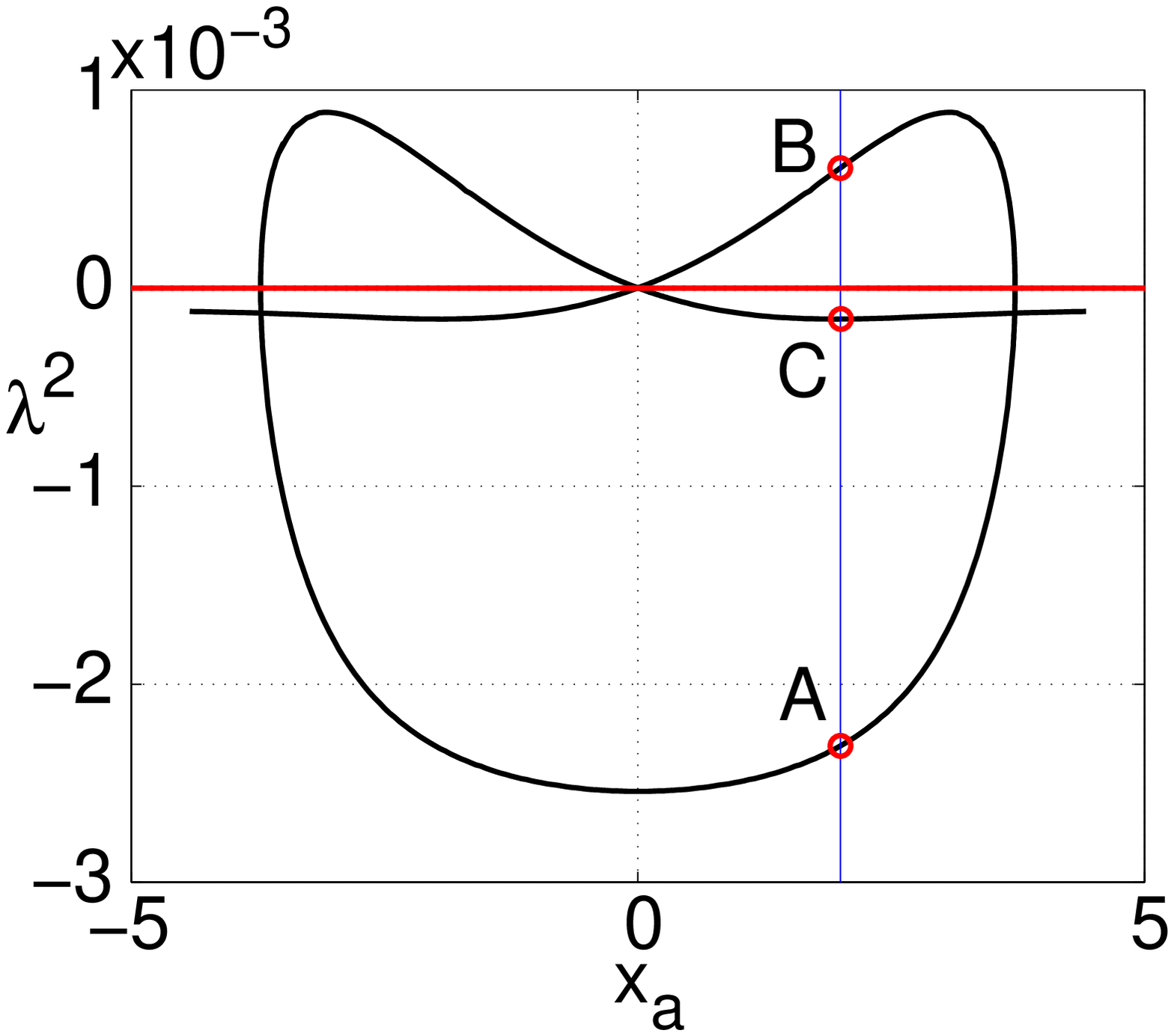}
   \includegraphics[width=4.5cm,angle=0]{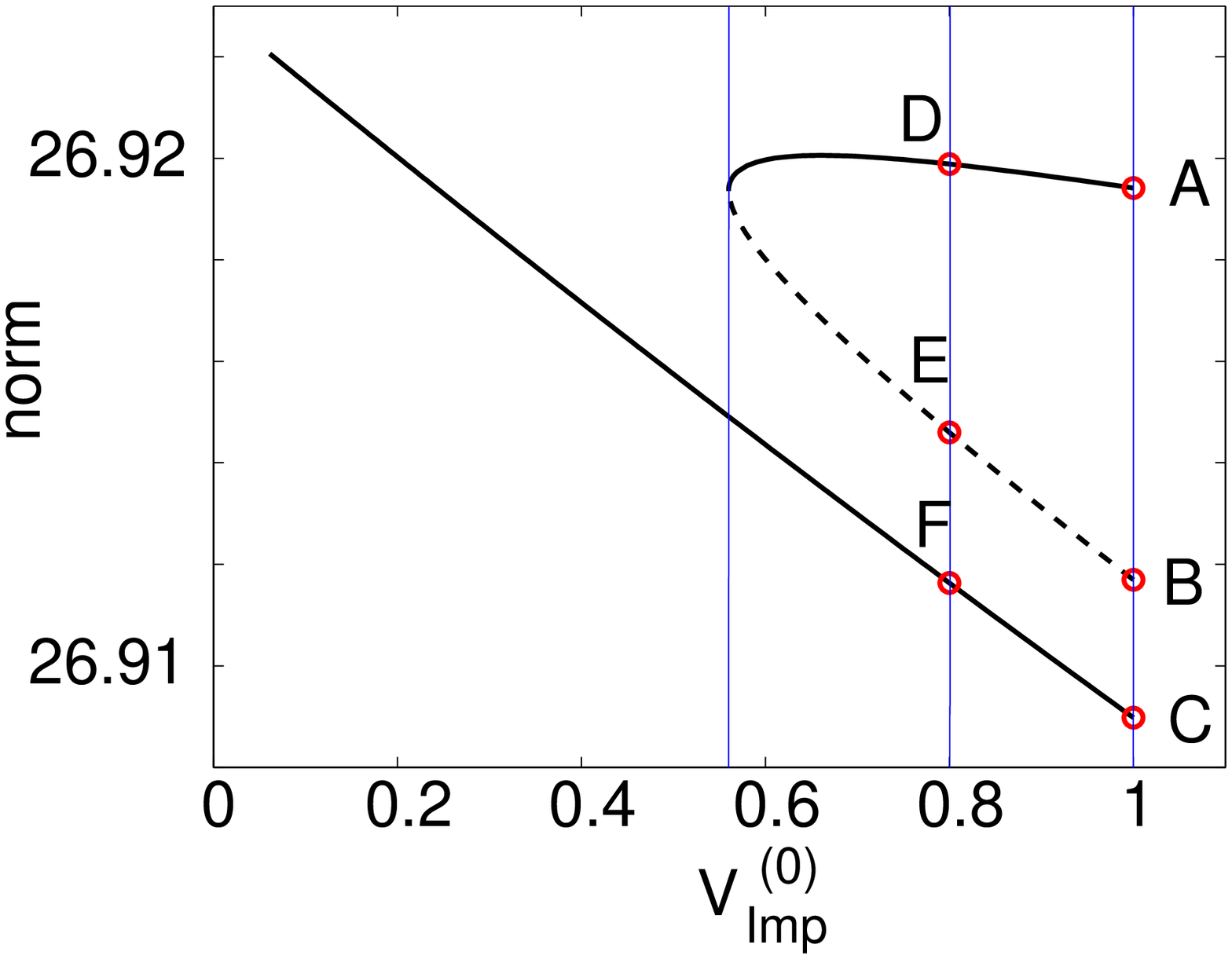}
%}
\caption{(Color online) The top panel is a one-dimensional slice
(for $y_a=0$) of the solution
surfaces (represented in the inset panel) as a function of the position of
the impurity $(x_a,y_a)$ for fixed $V_{\rm Imp}^{(0)}$.
The vertical axis corresponds to the $L^2$-norm squared of the solution
(i.e., %the square root of
the normalized number of atoms).
The dashed lines
(yellow (lighter) surface) correspond to unstable solutions, while the solid lines
(red (darker) surfaces) correspond to stable solutions.
The critical value of the radius, $x_{a,\rm{cr}}=3.72$ as well as the
characteristic value
$x_a=2$ and $x_a=0$ are represented
by vertical lines.
The bottom left panel shows the squared bifurcating
eigenvalues, $\lambda^2$, along these branches
($\lambda^2>0$ corresponds to an
instability), while the bottom right shows the saddle-node bifurcation as a
function of $V_{\rm Imp}^{(0)}$
for $x_a=2$ fixed (there are lines at
$V_{\rm Imp}^{\rm (cr)}=0.57$,
$V_{\rm Imp}^{(0)}=0.8$, and $V_{\rm Imp}^{(0)}=1$).
The solutions and spectra for each of the three branches,
represented by circles (and letters) for each of the characteristic
values of $x_a$ and $V_{\rm Imp}^{(0)}$ are presented in
Fig. \ref{stats}.
For these branches $(\mu,s,\varepsilon)=(1,1,0.5)$.
}
  \label{bif_rad}
\end{figure}

%%%%%%%%%%%%%%%%%%%%%%%%%%%%%%%%%%%%%%%%%%%%%%%%%%%%%%%%%%%%%%%%%%%%%%%%%%%%%

%%%%%%%%%%%%%%%%%%%%%%%%%%%%%%%%%%%%%%%%%%%%%%%%%%%%%%%%%%%%%%%%%%%%%%%%%
%
%\begin{figure}[ht]
%  %\newline
%  \centerline{
%    \includegraphics[width=7.cm,angle=0]{./norm_v0.eps}
%  }
%  \caption{(Color online) The bifurcation structure as a function
%of $V_0$ for $x_a=2$ fixed.  There are lines at $V_{0,\rm{cr}}=0.65$,
%$V_0=0.8$, and $V_0=1$.
%Notice the saddle node bifurcation at $V_{0,\rm{cr}}$ and the
%intersection of the two stable and one unstable branches with the solutions
%(d,e,f) presented in Fig. \ref{bif_rad}.  The solutions (g,h,i) are
%presented in Fig. \ref{v_stats}.
%}
%  \label{bif_v}
%\end{figure}
%
%%%%%%%%%%%%%%%%%%%%%%%%%%%%%%%%%%%%%%%%%%%%%%%%%%%%%%%%%%%%%%%%%%%%%%%%%%%%%

In this section we elaborate our investigation of the pinning
statics and the associated dynamical stability picture. In particular, we
thoroughly analyze the bifurcation structure of the
steady states with single-charge vorticity in the setting investigated
above (i.e. solutions to Eq.~(\ref{steadyst})) including their stability.
The latter will be examined by the eigenvalues of the linearization
around the steady state.
Upon obtaining a steady state solution $\Psi$ of Eq.~(\ref{steadyst}) and
considering a separable complex valued perturbation
$\tilde{u}=a(x,y) e^{\lambda t} + b^*(x,y) e^{\lambda^*t}$
of the steady state,
we arrive at the following eigenvalue problem for the growth rate,
$\lambda$, of the perturbation:

\begin{center}
\begin{math}
\bordermatrix{& \cr &\mathcal{L}_1 &\mathcal{L}_2 \cr
 &-\mathcal{L}_2^* &-\mathcal{L}_1^* \cr}
\bordermatrix{& \cr & a \cr & b \cr} = i \lambda \bordermatrix{& \cr & a \cr & b \cr},
\end{math}
\end{center}
where
\begin{eqnarray*}
\mathcal{L}_1 &=& -\mu-\frac{1}{2}(\partial_{x}^2+\partial_y^2)+V+2|\Psi|^2\\
\mathcal{L}_2 &=& \Psi^2.
\end{eqnarray*}

%, which will precess
%around the impurity under dynamical evolution,
%as shown in Fig. \ref{man_18}.
%See Fig. \ref{bif_scheme} for
%a 1-dimensional slice with $y=0$
%of
%the schematic of the bifurcation structure, represented by the potential
%$V$ for fixed $V^{(0)}_{\rm{Imp}}$ (in this section we will drop the
%unnecessary Imp and henceforth just use $V_0$) and varying the radius,
%$x_a$ in each of the three panels.  The three red dots in each panel
%represent the location of a vortex for a given solution at the given $x_a$.
%A cartoon of a slice of the potential
%with $y=0$ is shown for fixed $V^{(0)}_{\rm{Imp}}$ as the radius
%of the impurity offset, $x_a$ varies between zero (left panel)
%and $x_{a,\rm{cr}}$ (right panel) -
%the critical radius for pinning (shown by a red line).  The
%critical points of the potential  are indicated by red circles and
%represent the location of the vortex in the corresponding
%stationary vortex state for given $x_a$.
%The additional symmetry for $x_a=0$ implies
%that at the two minima of the potential the solutions are
%rotated versions of the same solution
%due to the additional symmetry
%as opposed to only $\theta \in \{0,1\}$ as in the
%diagram). There are three solutions for any given radius between
%(and not including) zero and $x_{a,\rm{cr}}$.

%%%%%%%%%%%%%%%%%%%%%%%%%%%%%%%%%%%%%%%%%%%%%%%%%%%%%%%%%%%%%%%%%%%%%%%%%

\begin{figure}[ht]
\begin{tabular}{c}
%\newline
 \includegraphics[width=5.0cm,angle=0]{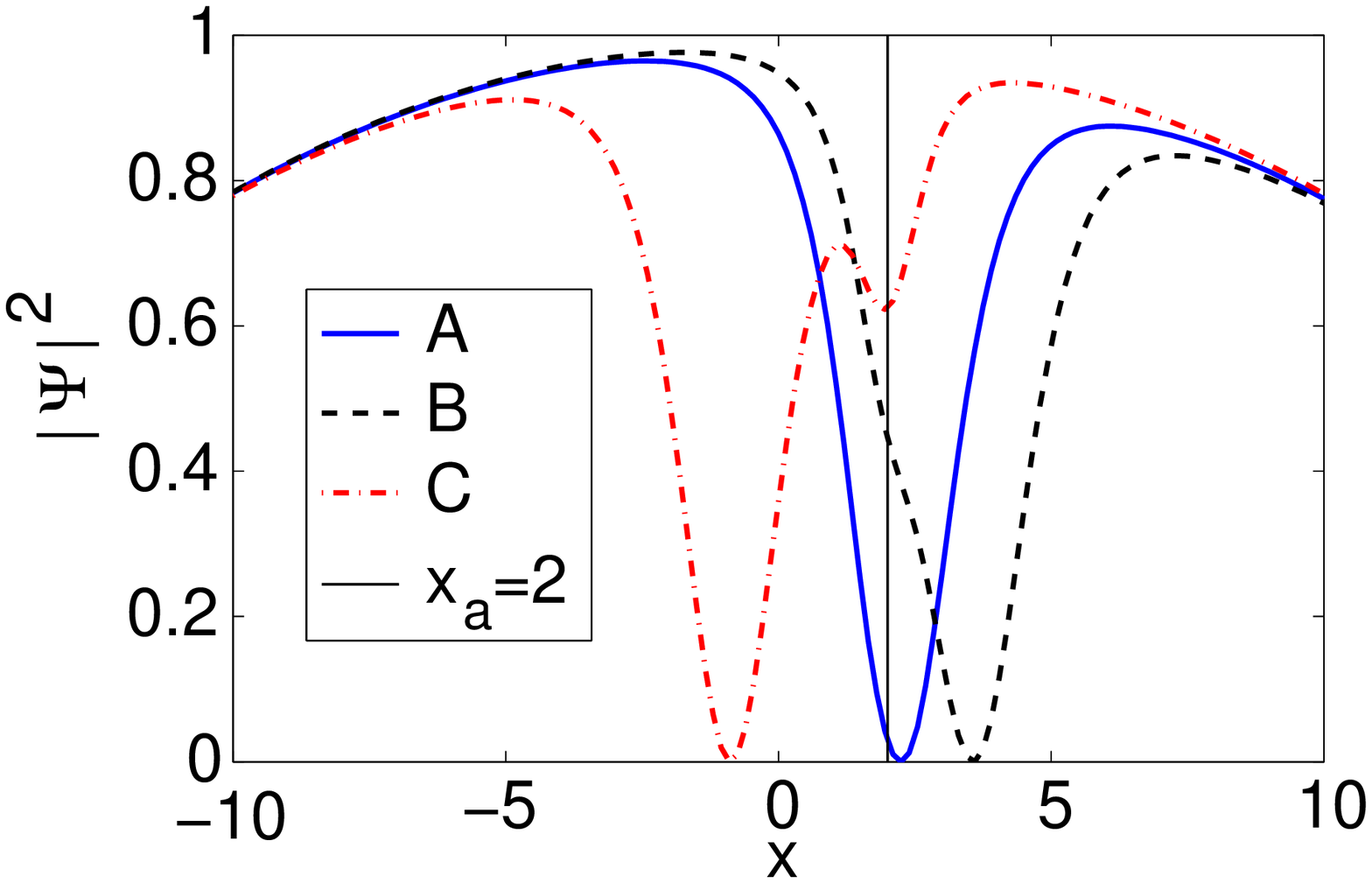}
\hskip-0.2cm
 \includegraphics[width=4.0cm,angle=0]{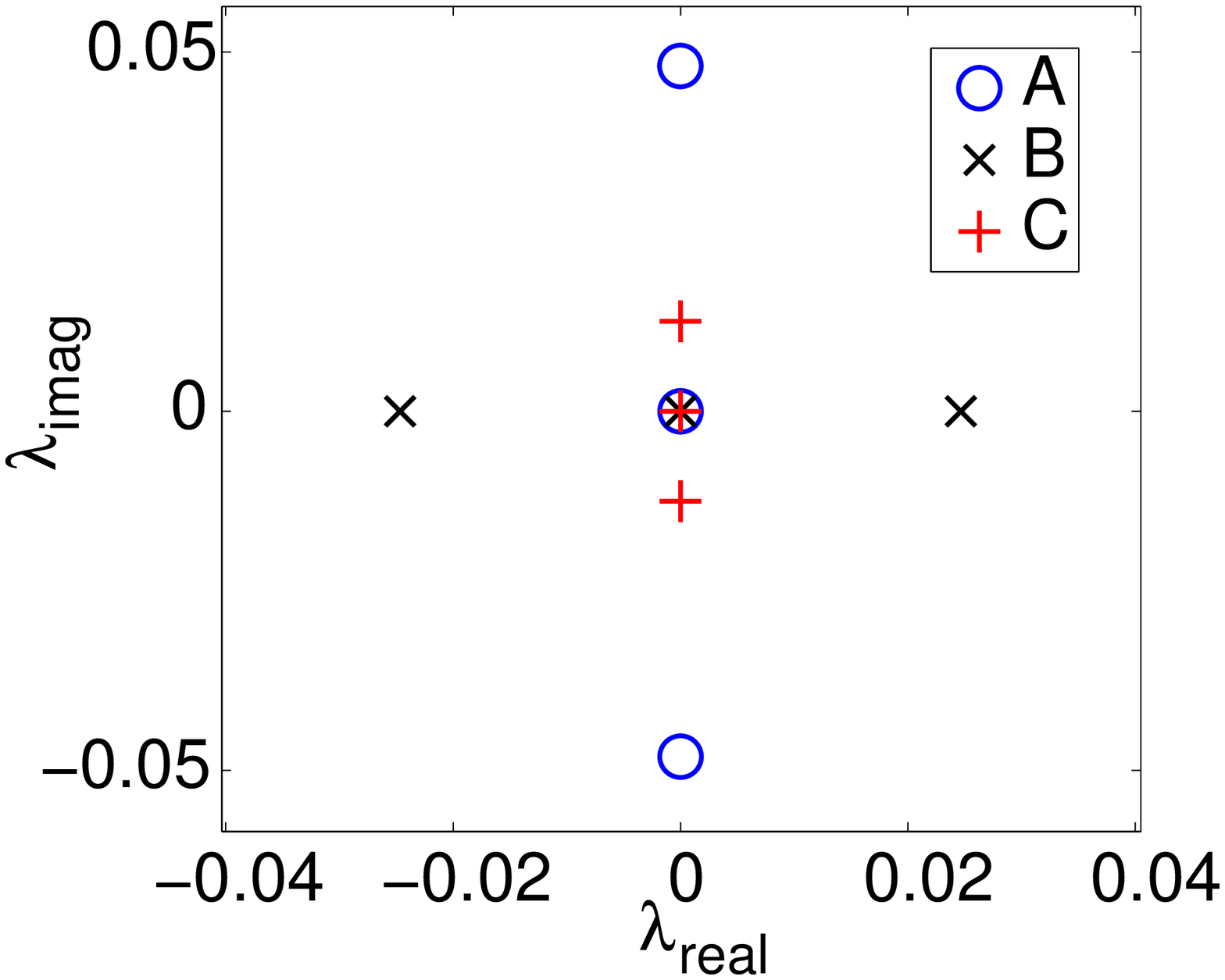}\\
 \includegraphics[width=5.0cm,angle=0]{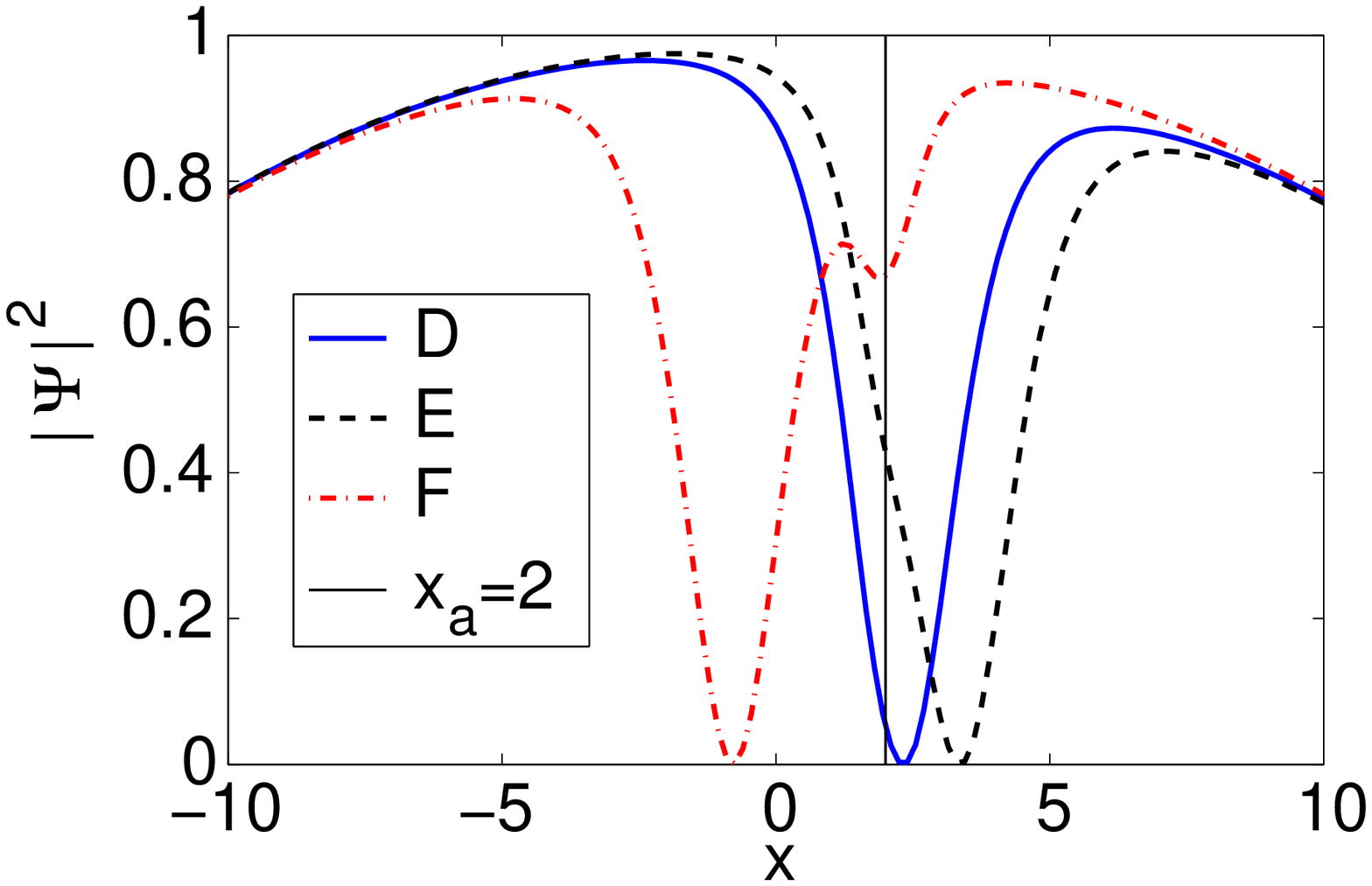}
\hskip-0.2cm
 \includegraphics[width=4.0cm,angle=0]{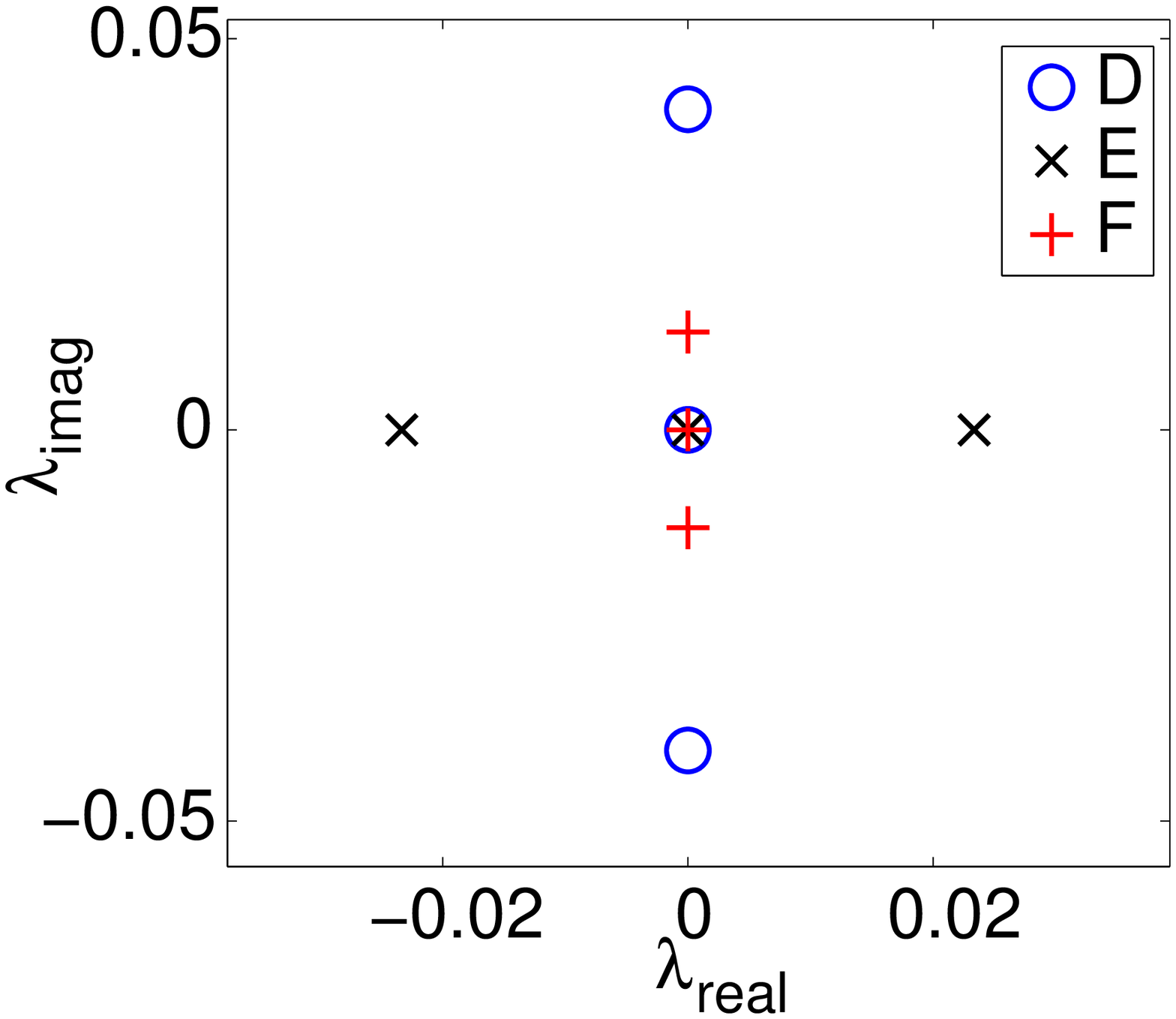}
\end{tabular}
  \caption{(Color online) The solutions (left, zoomed in to the region
where the vortices live) and corresponding
linearization spectra in a neighborhood of the origin (right),
for the various parameter values indicated by the circles
in Fig.~\ref{bif_rad}. The top (bottom) row corresponds to
$V_{\rm Imp}^{(0)}=1$ ($V_{\rm Imp}^{(0)}=0.8$).}
  \label{stats}
\end{figure}

%%%%%%%%%%%%%%%%%%%%%%%%%%%%%%%%%%%%%%%%%%%%%%%%%%%%%%%%%%%%%%%%%%%%%%%%%%%%%

%%%%%%%%%%%%%%%%%%%%%%%%%%%%%%%%%%%%%%%%%%%%%%%%%%%%%%%%%%%%%%%%%%%%%%%%%
%
%\begin{figure}[ht]
%  %\newline
%  %\centerline{
%\begin{tabular}{c}
%    \includegraphics[width=3.cm,angle=0]{./g_u.eps}
%  \includegraphics[width=3.cm,angle=0]{./g_eigen.eps}\\
%%\newline
%    \includegraphics[width=3.cm,angle=0]{./h_u.eps}
%  \includegraphics[width=3.cm,angle=0]{./h_eigen.eps}\\
%%\newline
%      \includegraphics[width=3.cm,angle=0]{./i_u.eps}
%  \includegraphics[width=3.cm,angle=0]{./i_eigen.eps}
%  %}
%\end{tabular}
%  \caption{(Color online) The same images as in Fig. \ref{rad_stats}
%except for the solutions labeled (g,h,i) in Fig. \ref{bif_v}.}
%  \label{v_stats}
%\end{figure}
%
%%%%%%%%%%%%%%%%%%%%%%%%%%%%%%%%%%%%%%%%%%%%%%%%%%%%%%%%%%%%%%%%%%%%%%%%%%%%%

%The critical points described by  represent
%saddle-node bifurcations in which the stable
%branch of
%solutions with the vortex pinned at the impurity
%collide with the respective unstable
%branch of the
%nearby steady state on the side of the impurity opposite the center of the
%harmonic trap.

As evidenced by the bifurcation diagrams presented in
Fig.~\ref{bif_rad},
there exist three solutions (i.e., steady-state
vortex positions) for any impurity displacement
radius in the interval $(0,x_{a,\rm{cr}})$,
letting $y_a=0$ without any loss of generality.
At the low end of the interval (i.e., for an
impurity at the center of the trap, $(x_a,y_a)=(0,0)$) there
is a transcritical bifurcation where the left (i.e., ones
for negative $x_a$) and right (i.e., for positive $x_a$)
solutions collide (see below for further explanation)
and exchange stability.
At the other end of the interval (i.e., for an
impurity at $(x_a,y_a)=(x_{a,\rm{cr}},0)$)
there exists a saddle-node bifurcation where two
new steady-state vortex solutions can be thought of
as emerging as $x_a$ decreases to values $x_a < x_{a,\rm{cr}}$,
or conversely can be thought of as disappearing as $x_a$ increases
to values $x_a > x_{a,\rm{cr}}$.
Among them, one
stable vortex position is found very close to the impurity
(see cases ``A'' and ``D'' in Figs.~\ref{bif_rad} and Fig.~\ref{stats}),
and another unstable vortex position further away
from the trap center
(see cases ``B'' and ``E'') in Figs.~\ref{bif_rad} and Fig.~\ref{stats}).
Considering only $x_a$, with $y_a=0$, two solution branches are
stable: one with the vortex sitting very close to the impurity
(see cases ``A'' and ``D'' in Figs.~\ref{bif_rad} and Fig.~\ref{stats}),
and one with the
vortex close to the center of the harmonic trap
(see cases ``C'' and ``F'' in Figs.~\ref{bif_rad} and Fig.~\ref{stats}).
The other solution is unstable
(see cases ``B'' and ``E'' in Figs.~\ref{bif_rad} and Fig.~\ref{stats}),
with the vortex sitting in the small effective potential minimum
on the side of the impurity
opposite the center of the harmonic trap.
This branch of solutions
collides with the one that has a vortex pinned at the impurity
and the two disappear in a
saddle-node bifurcation as the attraction of the impurity
becomes too weak to ``hold'' the vortex (this bifurcation corresponds
to the curves of critical parameter values in Fig. \ref{pin_exist}).
At $x_a=0$, the potential becomes radially
symmetric
and the solution with the vortex on the outside of the impurity
becomes identical (up to rotation) to the one close to the origin.
Indeed, at this point there is a single one-parameter
family of invariant solutions
with the vortex equidistant from the origin
(i.e., for any angle in polar coordinates), in addition to the single
solution with the vortex centered at $(0,0)$.
%For all other parameter values, there are three one-parameter
%families of invariant solutions.
The solutions in this invariant family, not being radially symmetric
and being in a radially symmetric trap,
necessarily have an additional pair of zero eigenvalues in the
linearization spectrum
to account for the additional invariance, i.e.~they
%solutions
have
4 instead of 2 zero eigenvalues
(note that the
%However, notice that the
solution with the vortex in the center for $x_a=0$
only has a single pair of zero eigenvalues due to its radial symmetry).
%is the radially symmetric one with vortex in the center for $x_a=0$,
%since the symmetry of the solution cancels the rotational invariance.
For $x_a<0$ the branches
exchange roles (transcritical bifurcation)
as the previously imaginary pair of eigenvalues for the stable
$x_a>0$ branch
(see cases ``C'' and ``F'' in Figs.~\ref{bif_rad} and Fig.~\ref{stats})
emerges on the real axis and the pair of previously
real eigenvalues from the unstable $x_a>0$ branch
(see cases ``B'' and ``E'' in Figs.~\ref{bif_rad} and Fig.~\ref{stats})
emerges on the imaginary axis, i.e. the branches
exchange their stability properties,
becoming the reflected versions of one another;
see the bottom left panel of Fig.~\ref{bif_rad} and the right column
of Fig.~\ref{stats}.
%The mediator of this transition is the invariant family of solutions for
%$x_a=0$, when there are 4 eigenvalues at the origin, as opposed to
%the regular 2, which are always there because of phase invariance.
%
In summary, for an impurity that is strong enough and
close enough to the trap center (specifically, $0 \leq x_a  < x_{a,\rm{cr}}$,
where $x_{a,\rm{cr}}$ depends of the strength of the impurity)
it is possible to {\em stably} pin the vortex very close to the impurity.
However, if the impurity is too far away from the trap's
center (specifically, $x_a>x_{a,\rm{cr}}$) the vortex can no longer
be pinned by the impurity.
The top image of Fig.~\ref{bif_rad}
depicts the bifurcations via
the $L^2$-norm squared (i.e., the
%the square root of the
normalized number of atoms) of the solution as a function of the radius, $x_a$
(for $y_a=0$),
%and as a
%radially symmetric two-parameter family of
%solutions (inset)
and as a function of arbitrary impurity location $(x_a,y_a)$ (inset).

In fact, the above picture holds for any fixed, sufficiently
large $V_{\rm Imp}^{(0)}$.
Conversely,
the same bifurcation structure can be represented by a continuation in the
amplitude of the impurity,
$V_{\rm Imp}^{(0)}$
(see bottom right panel of Fig. \ref{bif_rad}).
In particular, for a fixed radius of
$x_a=2$, a continuation was performed and a saddle-node bifurcation appears
for
$V_{\rm Imp}^{\rm (0,cr)} \approx 0.57$
where an unstable and a stable branch
emerge (corresponding exactly to those presented
in the continuation in $x_a$).
%For
%continuity, these branches are continued until the corresponding values of
%solutions (A,B,C) on the $x_a$ branches.
One can then infer that
the critical $V_{\rm Imp}^{(0)}$
decreases as $x_a$ decreases.
%, and, since there is
%only a single solution with the vortex sitting at the origin for
%$V_{\rm Imp}^{(0)}=0$ by interpolating
%in a full three-parameter space,
%the transcritical bifurcation must extend to a line that ends at the
%the 1D circular saddle-node bifurcation must necessarily extend to
%the origin where it meets
%a two-dimensional conic bifurcation, which meets the one-dimensional
%(in three-parameter space)
%transcritical bifurcation line
%at the origin and they both disappear.
%We note that one can also infer a similar cone for fixed
%$V_{\rm Imp}^{(0)}$
%and varying the width,
%$\varepsilon$, to zero.
%that is an
%extension of the circle of saddle-nodes!
%The solutions for (D,E,F) are presented in Fig. \ref{stats}.

%which can be represented, for instance, in radial coordinates with
%$\theta \in [0,2\pi)$

%%%%%%%%%%%%%%%%%%%%%%%%%%%%%%%%%%%%%%%%%%%%%%%%%%%%%%%%%%%%%%%%%%%%%%%%
\begin{figure}[ht]
  %\newline
  \centerline{
    \includegraphics[width=9.cm,height=7cm,clip]{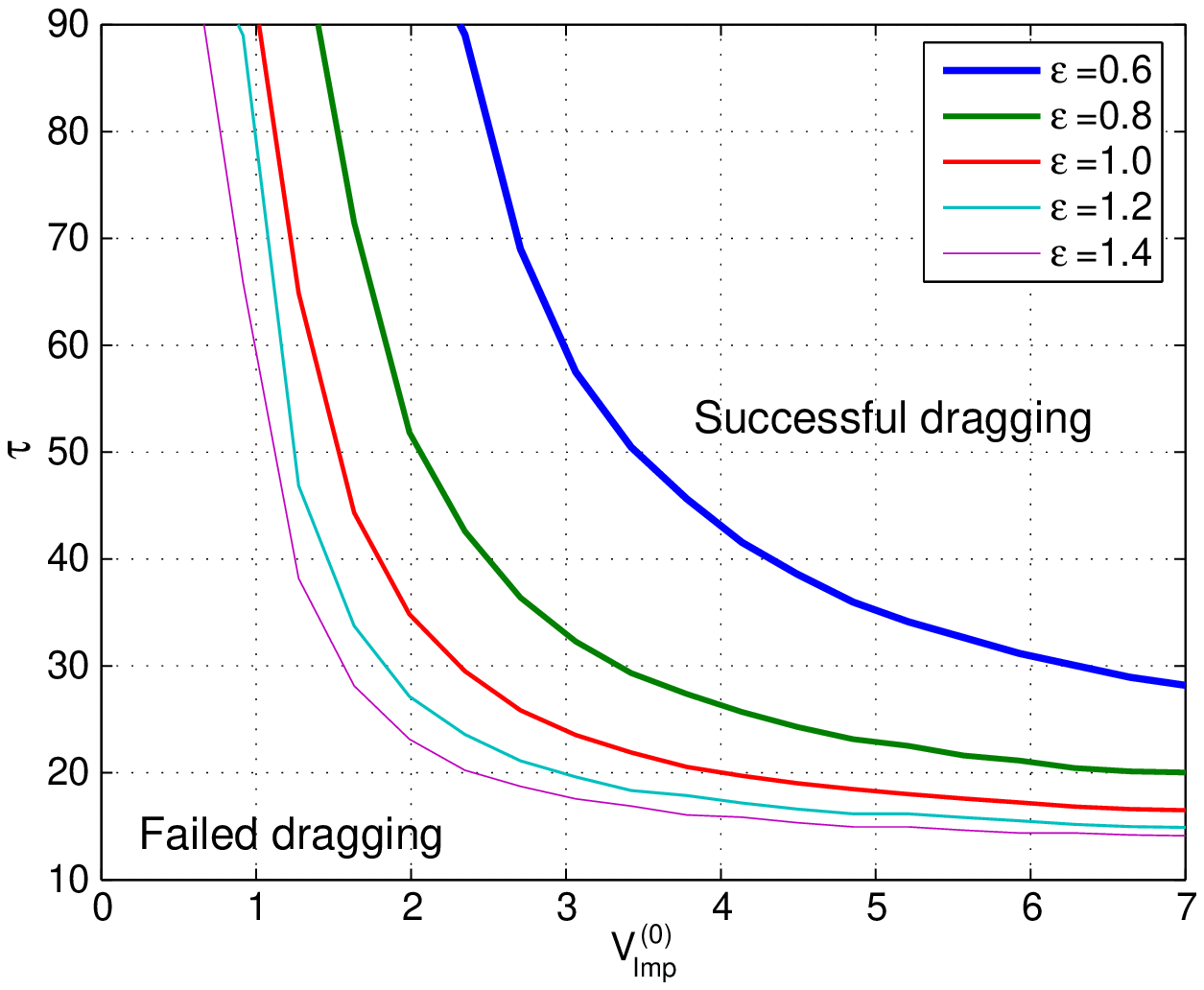}
  }
  \medskip
  \centerline{
    \includegraphics[width=9.cm,height=7cm,clip]{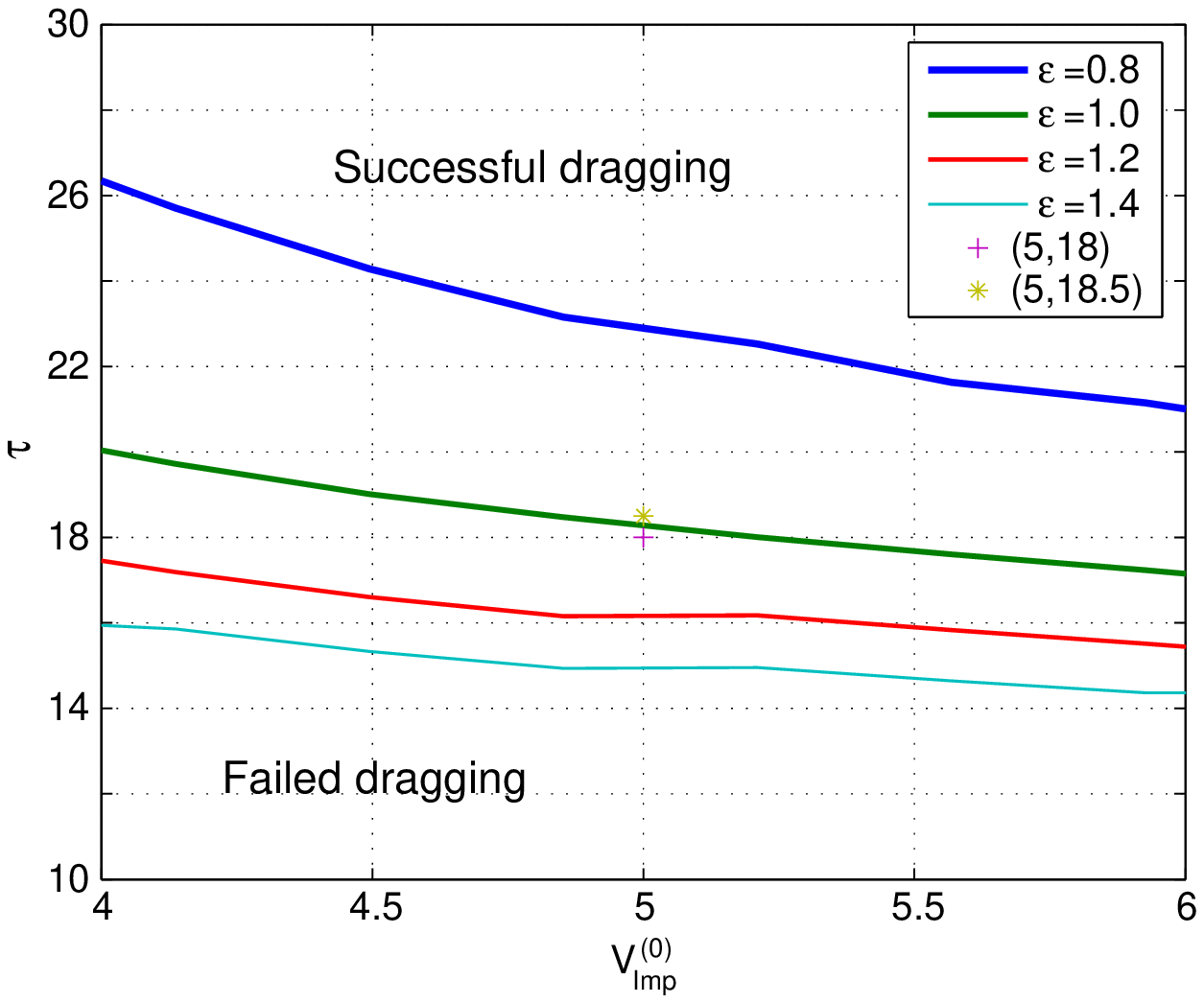}
  }
  \caption{(Color online) Parameter regions for successful manipulation of a
    single vortex inside a harmonic trap. The area above each
    curve corresponds to the successful dragging region.
%    The plot is created with a bisection method
%    in the $\tau$ parameter space with a tolerance set to $0.25$.
    The bottom panel depicts a zoomed region of the
    top panel where the asterisk and cross correspond,
    respectively, to the manipulation success and failure
    depicted in Fig.~\ref{man}.
    These panels indicate that higher intensity beams, and
%narrower
broader
    beams, can successfully drag vortices over shorter timescales
    than weaker,
%broader
narrower beams.
  }
  \label{pd_fig}
\end{figure}
\begin{figure*}[ht!]
 \centerline{
   \includegraphics[width=17cm,clip]{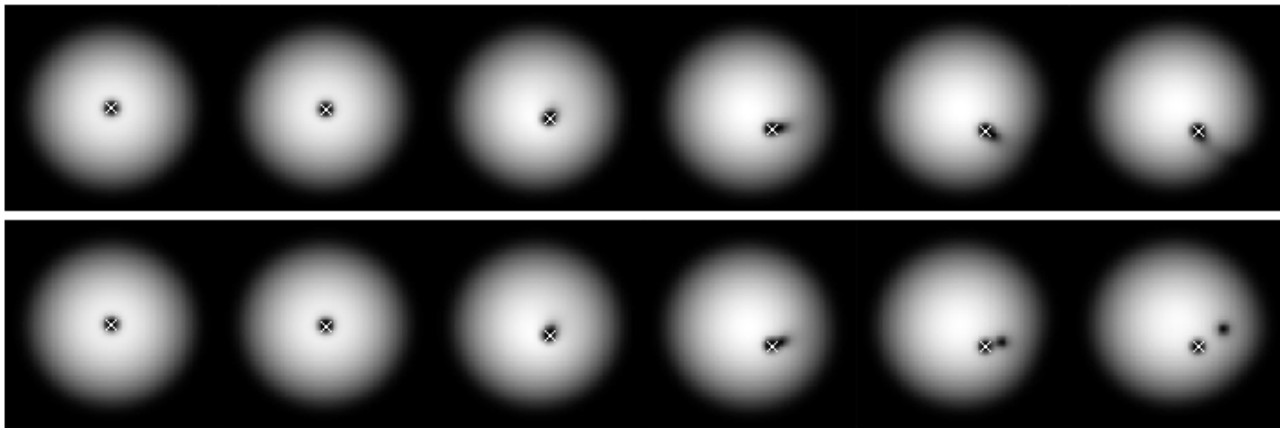}
 }
 \caption{Successful (top row) and failed (bottom row) vortex
   dragging cases by a moving laser impurity corresponding to the
   parameters depicted, respectively, by an asterisk and a cross
   in the bottom panel of Fig.~\ref{pd_fig}.
   In both cases, the laser impurity, marked by a cross, with
   $(V^{(0)}_{\rm Imp},\varepsilon)=(5,1)$ is moved adiabatically
   from $(0,0)$ to  $(5.43,-5.43)$ and the snapshots of the
   density are shown every $0.5 t^*$.
   The top row corresponds to a successful manipulation
   for $\tau = 18.5$ while the bottom row depicts a failed
   dragging for a slightly lower adiabaticity of $\tau = 18$.
 }
 \label{man}
\end{figure*}
%%%%%%%%%%%%%%%%%%%%%%%%%%%%%%%%%%%%%%%%%%%%%%%%%%%%%%%%%%%%%%%%%%%%%%%%

%%%%%%%%%%%%%%%%%%%%%%%%%%%%%%%%%%%%%%%%%%%%%%%%%%%%%%%%%%%%%%%%%%%%%%%%
\begin{figure*}[ht!]
  %\newline
  \centerline{
    \includegraphics[width=4.6cm,angle=-90]{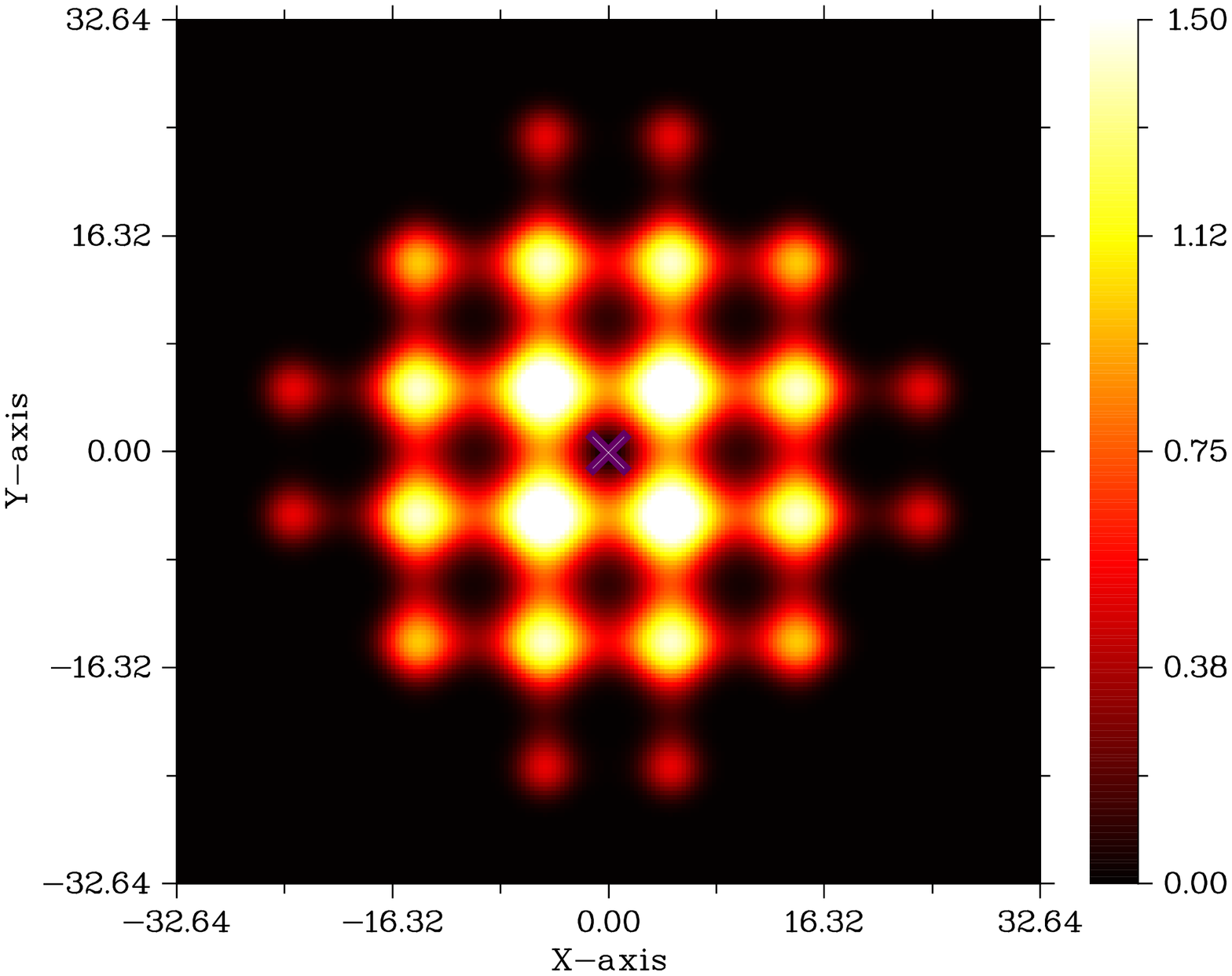}
    \includegraphics[width=4.6cm,angle=-90]{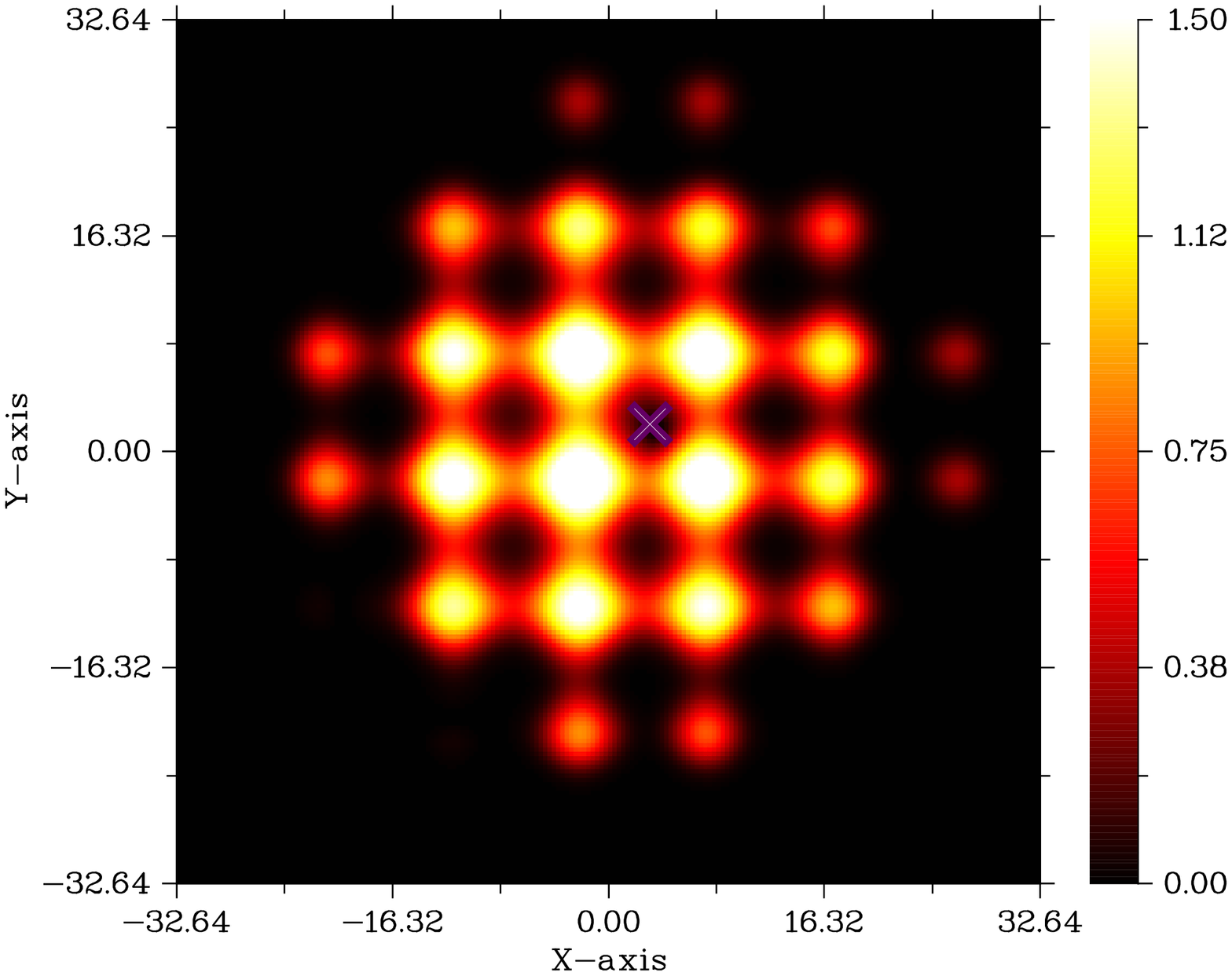}
    \includegraphics[width=4.6cm,angle=-90]{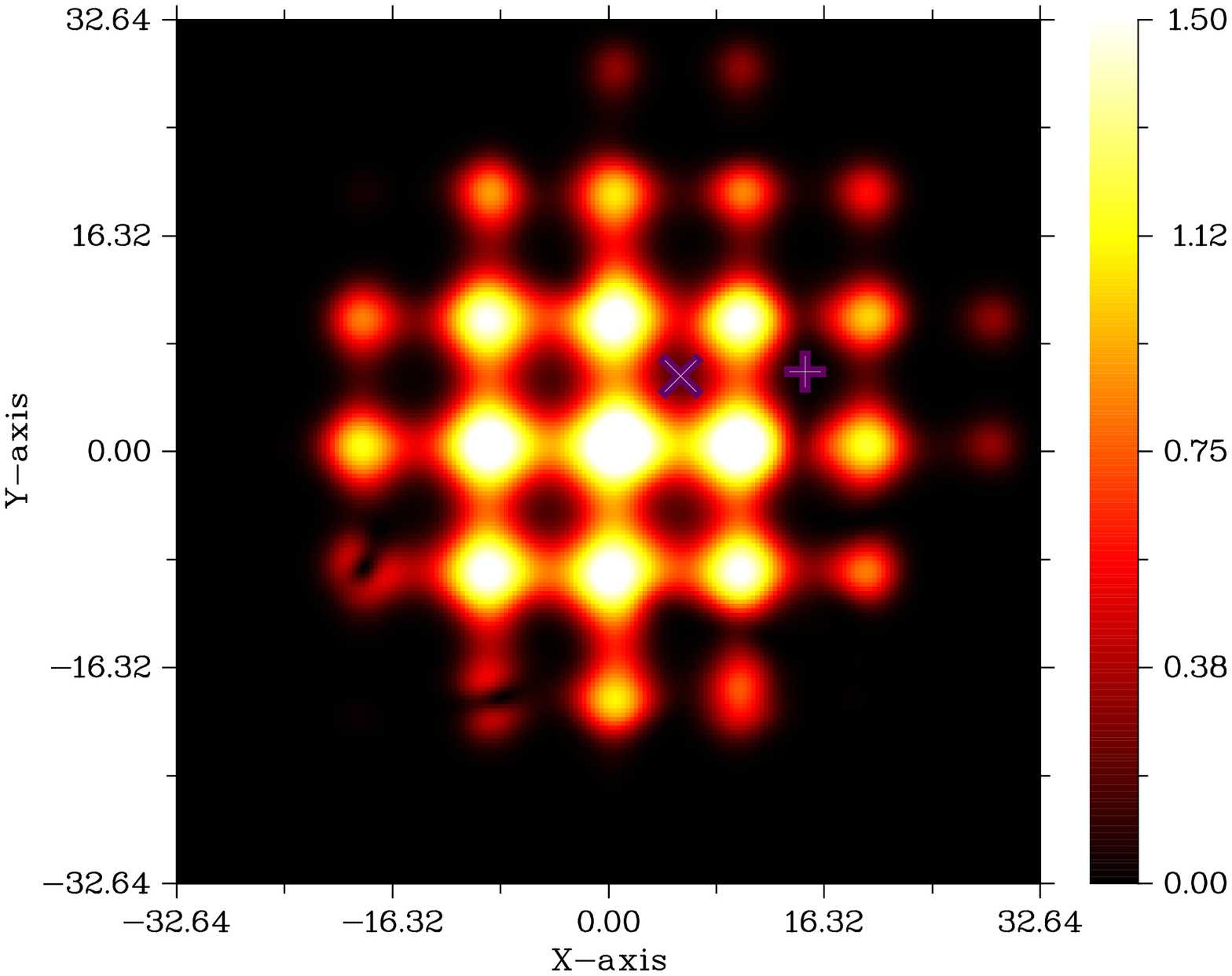}
  }
  \centerline{
    \includegraphics[width=4.6cm,angle=-90]{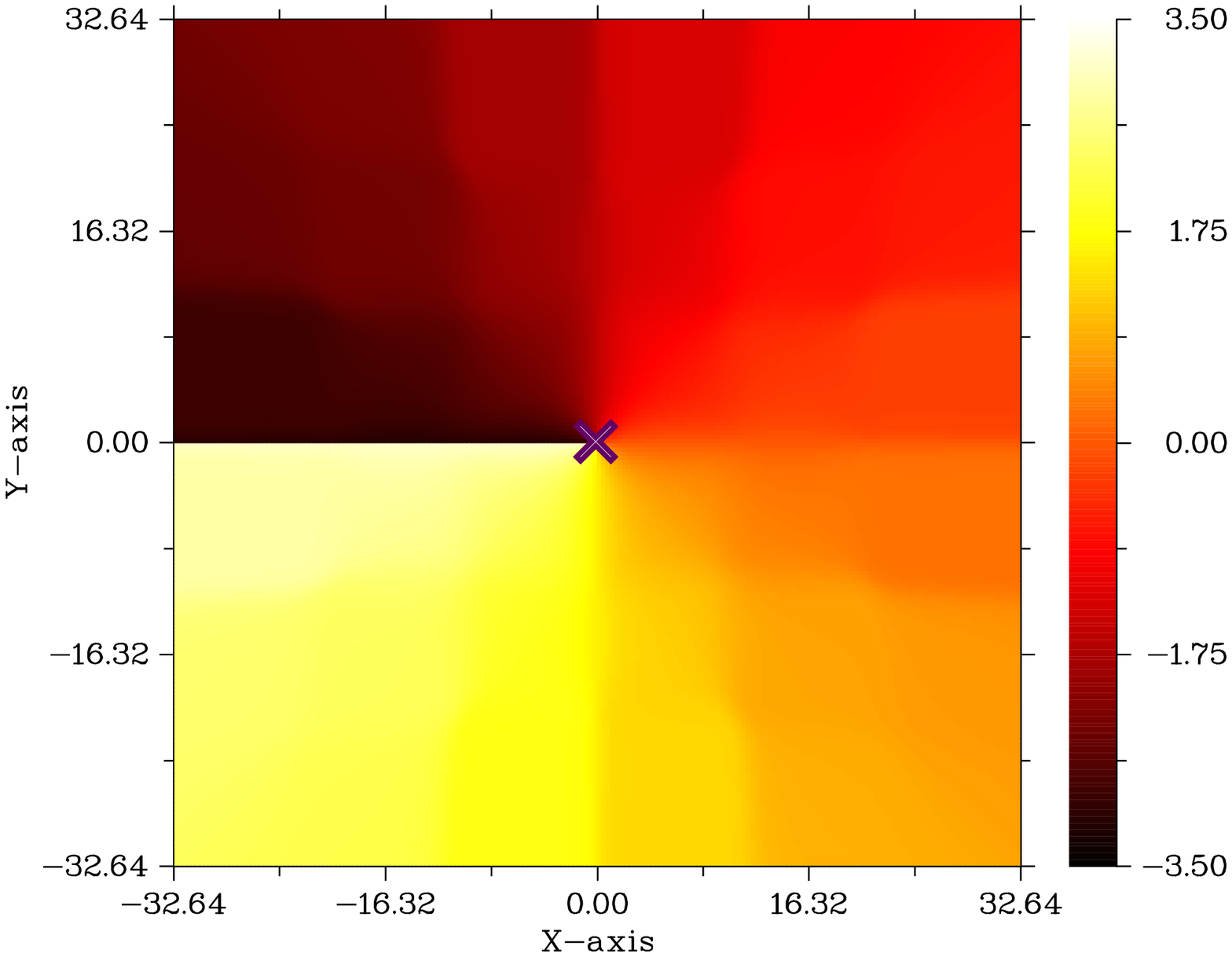}
    \includegraphics[width=4.6cm,angle=-90]{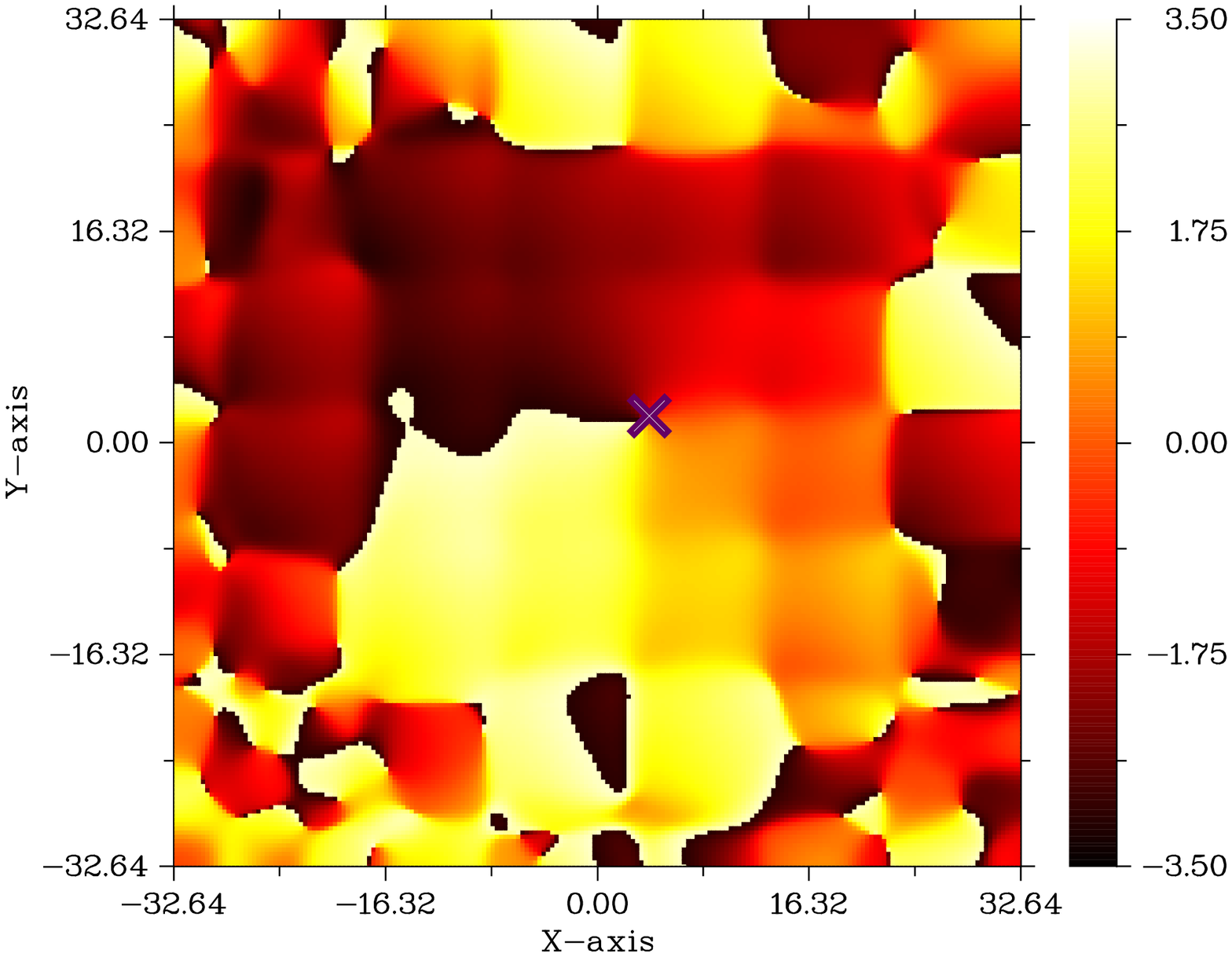}
    \includegraphics[width=4.6cm,angle=-90]{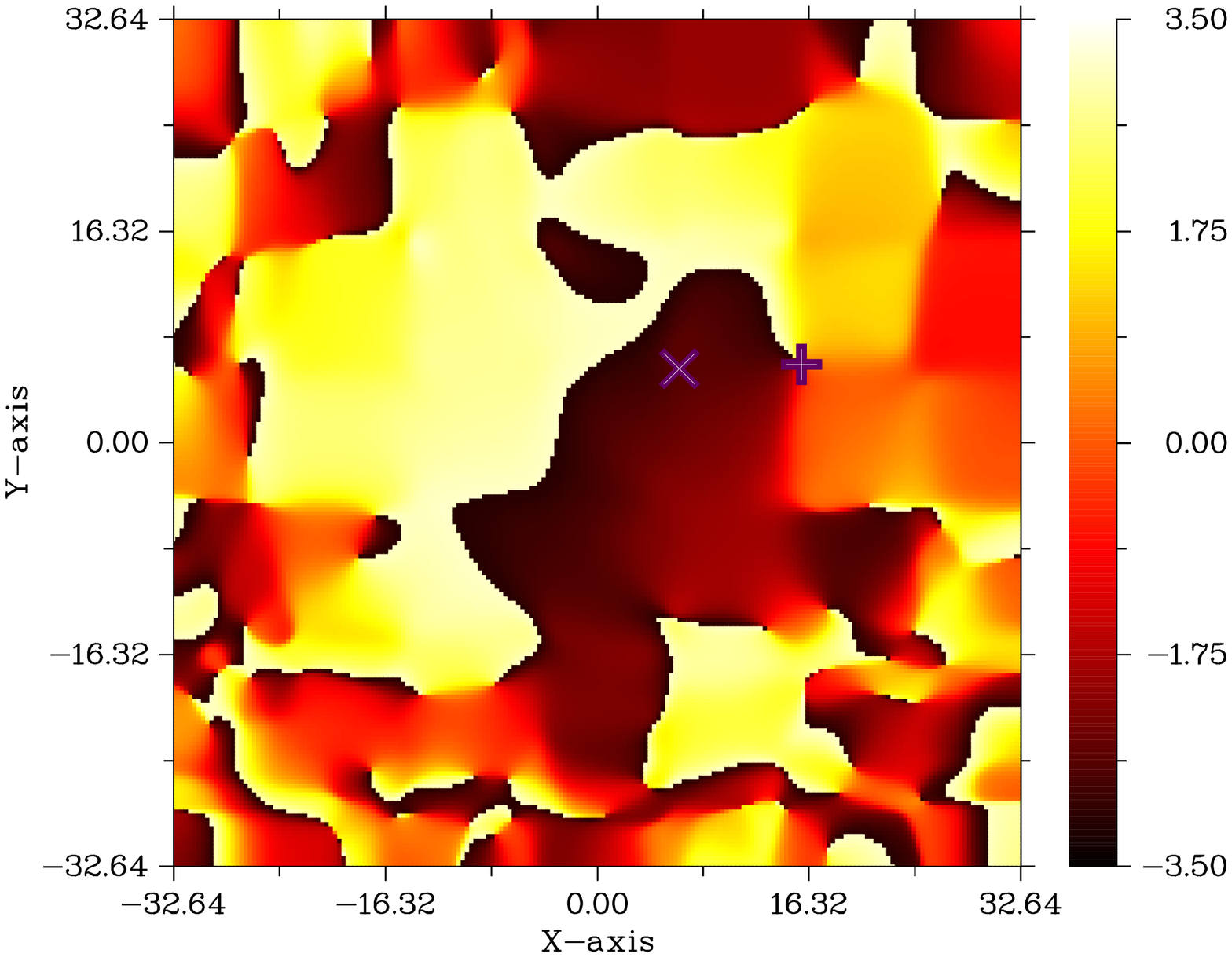}
  }
  \caption{(Color online)
Vortex dragging by an optical lattice potential
as in Eq.~(\ref{OL}). The central high-intensity maximum
of the optical lattice (depicted by a cross in the panels)
is moved adiabatically from the initial position $(x_a(0),y_a(0))=(0,0)$
to the final position $(5.43,5.43)$. The top row depicts
the BEC density (the colorbar shows the
density in adimensional units) while the bottom row depicts the phase of
the condensate where the vortex position (``plus'' symbol in
the right column) can be clearly inferred from the $2\pi$ phase
jump around its core. Observe how the vortex loses its
guiding well and jumps to a neighboring well in the
right column. The left, middle and right columns
correspond, respectively, to times $t=0$, $t=0.5t^*$ and
$t=2.5t^*$. The remaining parameters are as follows:
%$(V_{0},k,\tau,\theta_x,\theta_y) = (0.7,0.643,20,0,0)$.
$(V_{0},k,\tau,\theta_x,\theta_y) = (1.4,0.3215,20,0,0)$.
}
  \label{OLdrag}
\end{figure*}
%%%%%%%%%%%%%%%%%%%%%%%%%%%%%%%%%%%%%%%%%%%%%%%%%%%%%%%%%%%%%%%%%%%%%%%%

\section{The dynamical picture: Dragging and Capturing}\label{SEC:dragging}

\subsection{Vortex Dragging}

We would like now to take a pinned vortex and adiabatically drag it
with the impurity in a manner akin to what is has been
proposed for bright \cite{BS-Imp,BS-OL1,BS-OSL2} and dark
solitons \cite{DS-OL1,DS-OL2,DS-OL3}
in the quasi-1D configuration.
Manipulation of the vortex begins with the focused laser beam at the center of
the vortex. The laser is then adiabatically moved to a desired location
while continually tracking the position of the vortex.
Adiabaticity for the motion of the impurity is controlled by
the adiabaticity parameter $\tau$ controlling the acceleration
of the center $(x_a(t),y_a(t))$ of the impurity as:
\begin{eqnarray}
x_a(t) &=& x_{i} - \frac{1}{2}(x_{i}-x_{f})\left(1+
\tanh\left[\frac{t-t^*}{\tau}\right]\right),
\notag
\\[1.0ex]
y_a(t) &=& y_{i} - \frac{1}{2}(y_{i}-y_{f})\left(1+
\tanh\left[\frac{t-t^*}{\tau}\right]\right),
\end{eqnarray}
where the initial and final positions of the
impurity are, respectively, $(x_{i},y_{i})$ and
$(x_{f},y_{f})$. We will assume $y_a(t)=0$
(i.e., $y_i=y_f=0$) for the discussion below.
The instant of maximum acceleration is
\begin{equation}
t^* = \tanh^{-1}\left(\sqrt{1-\delta\tau}\right)\tau,
\end{equation}
where $\delta$ is a small parameter, $\delta = 0.001$, such that
the initial velocity of the impurity is negligible and
that $x_a(0) \approx x_{i}$ and $x_a(2t^*) \approx x_{f}$
(and the same for $y$).
This condition
on $t^*$ allows for the reduction of parameters
and allows us to ensure that we begin with a
localized impurity very close to the
center of the trap [i.e., $(x_{a}(0),y_{a}(0))\approx (0,0)$]
and that we will
drag it adiabatically to $(x_f,y_f)$ during the
time interval $[0,2t^*]$.
The next objective is to determine the relation between adiabaticity and
the various parameters such as strength
($V^{(0)}_{\rm Imp}$) or the width ($\varepsilon$) of the impurity in
order to successfully drag a vortex outward to a specific distance
from the center of the harmonic trap. In our study we set this distance
to be half of the radius of the cloud (half of the Thomas-Fermi radius).
We use the value $t^*$ to also define when to stop dynamically evolving
our system. In particular, we opt to continue monitoring
the system's evolution until $t_{f} = 3 t^*$.
%So that, as we change the adiabatic parameter $\tau$,
%our integrator compensates and continues integration until $t_{f} = 3 t^*$.
%
This choice ensures that a vortex that might have
been lingering
close to the impurity at earlier times would have either
been ``swallowed up'' by
the impurity and remain pinned for later times, or will have drifted further away due to the precession induced by the trap.

Applying this technique, along with a bisection method
(successively dividing the parameter step in half and changing
the sign of the parameter stepping once the
threshold pinning value is reached) within the span of relevant parameters
yields the phase diagram depicted in Fig.~\ref{pd_fig}.
The various curves in the figure represent the parameter boundaries
for successful dragging of the vortices for
different impurity widths (increasing widths from top to bottom).
All the curves for different widths are qualitatively similar
corresponding to
higher values of the adiabaticity parameter as the width is decreased.
This trend continues as $\varepsilon$ approaches the
existence threshold established in Fig.~\ref{pin_exist}.
In Fig.~\ref{man} we depict snapshots for the two cases
depicted by an asterisk (successful dragging) and a cross
(failed dragging) in the lower panel of Fig.~\ref{pd_fig}.

All of the numerical simulations discussed above deal with dragging
the vortex by means of the localized impurity. As with previous works of
vortex manipulations we also attempted to produce similar results
via an optical lattice (OL) potential generated by counter-propagating
laser beams \cite{BECBOOK}. In one dimension, the case of bright
solitons manipulated by OLs has been studied in
Refs.~\cite{BS-OL1,BS-OSL2} while the dark soliton case has
been treated in Refs.~\cite{DS-OL1,DS-OL2,DS-OL3}.
For a 1D OL, simply described by
$V^{1D}_{\rm OL}(x) = V_{0}\cos^2(kx+\theta_x)$ where
$k_x$ and $\theta_x$ are the wavenumber and phase of the OL,
the potential minima (or maxima) are isolated from each other
providing good effective potential minima for pinning and dragging.
On the other hand, when expanded to 2D, the OL reads:
\begin{equation}
V^{2D}_{\rm OL}(x,y) =
V_{0}\left[\cos^{2}(kx_a(t)+\theta_x)+\cos^{2}(ky_a(t)+\theta_y)\right],
\label{OL}
\end{equation}
where $k$ and $\theta_{x,y}$ are, respectively, the wavenumber
and phase of the OL in the $x$ and $y$ direction.
Here we observe that each 2D minimum (or maximum) is no longer
isolated, and that between two minima (or maxima) there are
areas for which the vortex can escape (near the saddle points of
the potential).
This is exactly what we observed when attempting to drag a vortex using
the 2D OL (\ref{OL}) without sufficient adiabaticity. The vortex would
meander around the various facets of
the lattice outside of our control.
To overcome this one needs to displace the potential with
a high degree of adiabaticity.
In doing so, we were successful in
dragging the vortices under some restraints
(relatively small displacements from the trap center).
An example of a partially successful vortex
dragging by an OL with potential
(\ref{OL}) is presented in Fig.~\ref{OLdrag}.
As it can be observed from the figure, the vortex (whose
center is depicted by a ``plus'') is dragged by the OL
(whose center is depicted by a cross) for some time.
However, before the OL reaches its final destination, the
pinning is lost and the vortex jumps to the neighboring
OL well to the right.
This clearly shows that vortex dragging with an OL is
a delicate issue due to the saddle points of the OL
that allow the vortex to escape.
Nonetheless, for sufficient adiabaticity, with a strong
enough OL and for small displacements from the trap center,
it is possible to successfully drag the vortex.
A more detailed study of the parameters that allow for
a successful dragging with the OL (i.e.,
relative strength and frequency of the lattice and
adiabaticity) falls outside of the scope
of the present manuscript and will be addressed in a future work.

%%%%%%%%%%%%%%%%%%%%%%%%%%%%%%%%%%%%%%%%%%%%%%%%%%%%%%%%%%%%%%%%%%%%%%%%%
%\begin{figure}[ht]
%  %\newline
%  \centerline{
%    \includegraphics[width=5.cm,height=7cm,angle=-90,clip]{./recap_ss.eps}
%  }
%  \medskip
%  \centerline{
%    \includegraphics[width=5.cm,height=7cm,angle=-90,clip]{./recap_ssp.eps}
%  }
%
%  \caption{(Color online) This figure shows a typical starting condition for the recapturing
%    simulation. Top: The density void on the left corresponds to the capturing
%    impurity and the one on the right is a vortex pinned by a second impurity.
%    When integration begins the pinning impurity is adiabatically turned off,
%    releasing the vortex into the precession of the trap. Bottom: Complex phase
%    plot illustrating that the figure to the right is the only vortex.
%  }
%  \label{ss_recap}
%\end{figure}
%%%%%%%%%%%%%%%%%%%%%%%%%%%%%%%%%%%%%%%%%%%%%%%%%%%%%%%%%%%%%%%%%%%%%%%%%

%%%%%%%%%%%%%%%%%%%%%%%%%%%%%%%%%%%%%%%%%%%%%%%%%%%%%%%%%%%%%%%%%%%%%%%%
\begin{figure}[ht]
  %\newline
  \centerline{
    \includegraphics[width=7.cm,angle=-90]{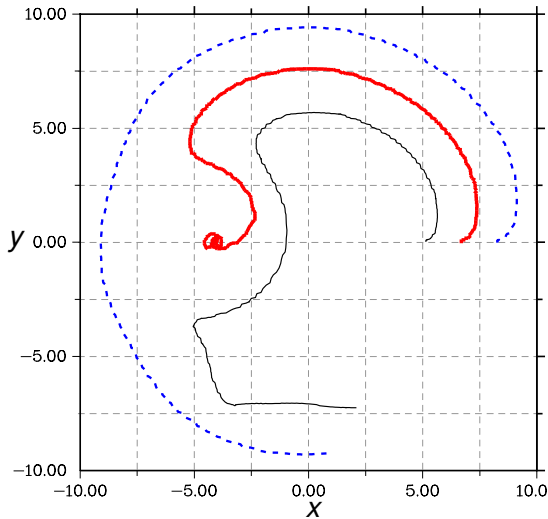}
  }
  \caption{(Color online)
Capturing a precessing vortex by a stationary impurity.
    The different paths correspond to isolated vortices that are
    released by adiabatically turning off a pinning impurity
    at the following off-center locations:
    $(5,0)$ (thin black line), $(6.5,0)$ (thick red line) and
    $(8,0)$ (blue dashed line). The capturing impurity is
    located at $(-4,0)$. The first and third cases fail to produce
    capturing while the second case manages to capture the vortex. One
    interesting feature that we observed in the case of successful capture is
    that, before its gets captured, the
    vortex gets drawn into the impurity potential, but then almost gets
    knocked back out by the phonon radiation waves created from the capture
    which bounce around within the condensate.}
  \label{recap}
\end{figure}
%%%%%%%%%%%%%%%%%%%%%%%%%%%%%%%%%%%%%%%%%%%%%%%%%%%%%%%%%%%%%%%%%%%%%%%%

\subsection{Vortex Capturing}

A natural extension of the above results is to investigate whether it is possible to
capture a vortex that is already precessing by an appropriately
located and crafted impurity.
This idea of capturing, paired with the dragging ability, suggests that a
vortex created off-center, which is typically the case in an experimental
setting, can be captured, pinned and dragged to a desired location either at
the center of the trap or at some other distance off-center.
We now give a few examples %where we were able
demonstrating that it is indeed possible for a localized impurity to capture a moving vortex.
The simulation begins with a steady-state solution of a
vortex pinned by an impurity at a prescribed radius and a second impurity
on the opposite side of the trap at a different radius. Initial numerical
experiments have been done to determine the importance of the difference in these
distances from the trap center.

As is shown in Fig.~\ref{recap} the capturing
impurity must be located sufficiently lower
(i.e., closer to the trap center) than the trapping impurity
in order for the vortex to be pulled from its precession and be
captured by the impurity. Intuitively one might come to the conclusion
that if the vortex and impurity were located the same distance away from
the center of the trap, then the
vortex should be captured. But due to the interaction between the vortex and
the impurity that was discussed
earlier, as the vortex approaches the impurity,
it begins to interact with it by %rotating
precessing
clockwise around
the impurity. Thus the orientation of the vortex and impurity with respect to the
trap center greatly determines the dynamics.
This combination of the interactions of the vortex with the trap and the vortex
with the impurity then dictates that for a vortex to be captured by the
impurity while precessing around the harmonic trap and rotating around the impurity,
the impurity must be positioned at least closer to the trap center than the initial distance between the vortex and the trap center.
%We are pursuing more detailed investigations of the process of vortex
%capturing in conjunction with building an experimental setup where
%these predictions can be tested and anticipate to report relevant
%observations in future publications.
%%for various more complicated conditions, and plan to test our predictions with experimental observations.

\section{Conclusions\label{SEC:conclu}}

In summary, we studied the effects on isolated vortices
of a localized impurity generated
by a narrowly focused laser beam
inside a parabolic potential in the context of
Bose-Einstein condensates (BECs). We not only examined the
dynamics (dragging and capture) of the vortex solutions in
this setting, but
also analyzed in detail the stationary (pinned vortex) states, their linear
stability and the
underlying bifurcation structure of the problem.

As is already well known, the harmonic trap is responsible
for the precession of the vortex around the condensed cloud.
We have further demonstrated that a narrowly focused blue-detuned
laser beam induces a local
attractive potential that is able to pin the vortex at
various positions within the BEC, and we investigated
the dependence of pinning
as a function of the laser beam parameters (width and power)
for different locations in the condensed cloud.
For a fixed beam width, we then explored the underlying bifurcation
structure of the stationary solutions in the parameter space
of pinning position and beam power.
We found that for sufficiently high beam intensity it
is possible to overcome the vortex precession
and to {\em stably} pin the vortex at a desired
position inside the condensed cloud.
We also studied the conditions for a vortex to be
dragged by an adiabatically moving beam and concluded
that for sufficiently high intensity beams and for
sufficient adiabaticity it is possible to drag the
vortex to almost any desired position within the BEC cloud.
The possibility of vortex dragging using periodic, two-dimensional,
optical lattices was also briefly investigated.
Due to the lattice's saddle points between consecutive
wells, the vortex is prone to escape to neighboring
wells and, therefore, dragging with optical lattices
is arguably less robust that its counterpart with
focused laser beams.
Finally, we presented the possibility of capturing a precessing
vortex by a stationary laser beam. Due to the combined action of
the precession about the harmonic trap and the precession about
the localized impurity, the stationary laser must be
carefully positioned to account for both precessions
so that the vortex can be successfully captured by the laser beam.

%We identified a clear zero-dimensional
%transcritical bifurcation at the origin in impurity displacement
%parameter space, as well as a one-dimensional, radially symmetric
%(circle) saddle-node bifurcation
%(one point for an equal radius in any direction).  For a
%smaller power impurity, the
%saddle-node bifurcation was found to occur at a smaller radius,
%and hence, a conic surface bifurcation was identified which
%disappears along with the transcritical one at zero power
%and displacement radius.
%The saddle-node bifurcation is responsible for the
%critical radius for pinning.
%

%The pinning of the vortex by the localized beam opens the
%possibility of adiabatically dragging a vortex through a BEC by slowly varying
%the location of the laser beam.
%In our first investigations of this phenomenon, we found that a key parameter
%for successful dragging is the degree of adiabaticity of the laser motion
%(together with the width and power of the beam) and we
%described the parameter regions where dragging may be possible.
%
%We also briefly described the possibility of dragging the
%vortex by using a periodic optical lattices but we found
%that this dragging method is much less robust than the
%one with the localized laser beam.
%

This work paves the way for a considerable range of future studies
on the topic of vortex-impurity interactions.
Among the many interesting possibilities that can be considered,
we mention the case of more complex initial conditions, such as
higher topological charge ($\pm s$) vortices, and that of complex dynamics
induced by the effects of multiple laser beams.  For example, in the
latter setting, we might
envision a situation in which a single vortex is localized to a region
within a BEC by appropriate dynamical manipulation of multiple laser
beams \emph{without} relying on vortex pinning.  Such additional studies may
provide a more complete understanding of the physics of manipulating vortex
arrays by optical lattices.  Additional investigations will also need to
consider the role of finite temperature and damping, as well as the
consequences of moving impurities located near the Thomas-Fermi radius where
density is low and critical velocities for vortex shedding are much lower
than near the BEC center.

Another natural extension of our work is to study the manipulation
of vortex lines in three-dimensional condensates. It would be
interesting to test whether the beam could stabilize a whole
vortex line (suppression of the so-called
Kelvin modes \cite{Bretin2003a}) and, moreover, change the orientation of a vortex \cite{Haljan2001a}. Along this
vein, a more challenging problem would be to study the pinning
and manipulation of vortex rings by laser sheets; see e.g. \cite{hau}.
These settings would also present the possibility of identifying a richer
and higher-dimensional bifurcation structure.
%
%We are currently studying these avenues and the results
%will be reported in a future publication.

%Through the use of analytical and numerical techniques, we have shown
%particular parameters which are sufficient for the pinned, stable vortex
%solutions at a particular distance off-center to the HT.
%We then continued by establishing regions of
%parameter values for which vortices could be dragged from a steady state
%solution at the center of the HT to a predetermined distance away from the
%center utilizing the stability conditions as a lower bound for our
%simulations.

\vspace{5mm}

{\bf Acknowledgements}. PGK gratefully acknowledges support from
the NSF-CAREER program (NSF-DMS-0349023), from NSF-DMS-0806762
and from the Alexander von Humboldt Foundation. RCG gratefully
acknowledges support from NSF-DMS-0806762.


\begin{thebibliography}{10}

\bibitem{Anderson1962aCampbell1972aDaldini1974aCivale1991a}
%theory of flux creep in hard superconductors
P.W.\ Anderson,
Phys.\ Rev.\ Lett.\ {\bf 9}, 309 (1962);
%
% Flux vortices and transport currents in type II superconductors
A.M.\ Campbell and J.E.\ Evetts,
Advan.\ Phys.\ {\bf 21}, 199 (1972);
%
%vortex-line pinning by thickness modulation of superconducting films
O.\ Daldini, P.\ Martinoli, and J.L.\ Olsen,
Phys.\ Rev.\ Lett.\ {\bf 32}, 218 (1974);
%
%vortex confinement by columnar defects in YBa2Cu3O7 crystals:
% enhanced pinning at high fields and temperatures
L.\ Civale \emph{et al.},
Phys.\ Rev.\ Lett.\ {\bf 67}, 648 (1991).


\bibitem{Baert1995aReichhardt2001aGrigorenko2003a}
%composite flux line lattices stabilized in superconducting films by a regular array of artificial defects
M.\ Baert, V.V.\ Metlushko, R.\ Jonckheere, V.V.\ Moshchalkov, and Y.\ Bruynseraede,
Phys.\ Rev.\ Lett.\ {\bf 74}, 3269 (1995);
%
% commensurate and incommensurate vortex lattice melting in periodic pinning arrays
C.\ Reichhardt, C.J.\ Olson, R.T.\ Scalettar, and G.T.\ Zim\'{a}nyi,
Phys.\ Rev.\ B {\bf 64}, 144509 (2001);
%
% symmetry locking and commensurate vortex domain formation in periodic pinning arrays
A.N.\ Grigorenko \emph{et al.},
Phys.\ Rev.\ Lett.\ {\bf 90}, 237001 (2003).


\bibitem{Reijnders2004aPu2005aReijnders2005a}
% pinning of vortices in a bose-einstein condensate by an optical lattice
J.W.\ Reijnders and R.A.\ Duine,
Phys.\ Rev.\ Lett.\ {\bf 93}, 060401 (2004);
%
% structural phase transitions of vortex matter in an optical lattice
H.\ Pu, L.O.\ Baksmaty, S.\ Yi, and N.P.\ Bigelow,
Phys.\ Rev.\ Lett.\ {\bf 94}, 190401 (2005);
%
% pinning and collective modes of a vortex lattice in a bose-einstein condensate
J.W.\ Reijnders and R.A.\ Duine,
Phys.\ Rev.\ A {\bf 71}, 063607 (2005).



\bibitem{Tung2006a}
%observation of vortex pinning in bose-einstein condensates
S.\ Tung, V.\ Schweikhard, and E.A.\ Cornell,
Phys.\ Rev.\ Lett.\ {\bf 97}, 240402 (2006).


\bibitem{Bhat2006aGoldbaum2008a}
%bose-einstein condensates in rotating lattices
Rajiv Bhat, L.D.\ Carr, and M.J.\ Holland,
Phys.\ Rev.\ Lett.\ {\bf 96}, 060405 (2006);
%
%vortex lattices of bosons in deep rotating optical lattices
Daniel S.\ Goldbaum and Erich J.\ Mueller,
Phys.\ Rev.\ A {\bf 77}, 033629 (2008).


\bibitem{Dahl2008a}
%unusual states of vortex matter in mixtures of bose-einstein condensates on rotating optical lattices
E.K.\ Dahl, E.\ Babaev, and A.\ Sudb{\o},
Phys.\ Rev.\ Lett.\ {\bf 101}, 255301 (2008).


\bibitem{Geurts2008a}
%topologically trapped vortex molecules in bose-einstein condensates
R.\ Geurts, M.V.\ Milo\v{s}evi\'{c}, and F.M.\ Peeters,
Phys.\ Rev.\ A {\bf 78}, 053610 (2008).


\bibitem{BECBOOK}
P.G.~Kevrekidis, D.J.~Frantzeskakis, and R.~Carretero-Gonz\'alez (eds).
{\sl Emergent Nonlinear Phenomena in Bose-Einstein Condensates:
Theory and Experiment}.
Springer Series on Atomic, Optical, and Plasma Physics,
Vol.~{\bf 45}, 2008.

\bibitem{NonlinearityReview}
R.~Carretero-Gonz{\'a}lez, D.J. Frantzeskakis and P.G. Kevrekidis.
%Nonlinear waves in {B}ose-{E}instein condensates: Physical relevance
%  and mathematical techniques.
Nonlinearity, {\bf 21} R139 (2008).

\bibitem{Manipulation-SPIE}
% Optical Manipulation of Matter Waves.
R. Carretero-Gonz\'alez, P.G. Kevrekidis, D.J. Frantzeskakis, and B.A. Malomed.
Proc. SPIE Int. Soc. Opt. Eng. {\bf 5930} (2005) 59300L.

\bibitem{BS-Imp}
G. Herring, P.G. Kevrekidis, R.\ Carretero-Gonz\'alez, B.A. Malomed,
D.J. Frantzeskakis, and A.R. Bishop.
\newblock {\em Phys.\ Lett.\ A}, {\bf 345} (2005) 144.

\bibitem{BS-OL1}
%Statics, Dynamics and Manipulation of Bright Matter-wave Solitons in Optical Lattices.
P.G.\ Kevrekidis, D.J.\ Frantzeskakis, R.\ Carretero-Gonz\'alez, B.A.\ Malomed,
G.\ Herring, and A.R.\ Bishop.
\newblock {\em Phys.\ Rev.\ A}, {\bf 71} (2005) 023614.

\bibitem{BS-OSL2}
%\newblock {Statics, dynamics and manipulation of bright matter-wave solitons in
%  optical lattices}.
M.A.\ Porter, P.G.\ Kevrekidis, R.\ Carretero-Gonz\'alez, and D.J.\ Frantzeskakis.
\newblock {\em Phys.\ Lett.\ A},  {\bf 352} (2006) 210.

\bibitem{DS-Imp}
%Interaction of a dark soliton with a localized impurity
V.V.\ Konotop, V.M.\ P\'erez-Garc\'ia, Y.-F.\ Tang and L.\ V\'azquez,
Phys.\ Lett.\ A {\bf 236}, 314 (1997).

\bibitem{vvkve} V.V. Konotop and V.E. Vekslerchik,
Phys.\ Rev.\ E {\bf 49}, 2397 (1994).

\bibitem{fr1}
%Interaction of dark solitons with localized impurities in Bose-Einstein condensates
D.J. Frantzeskakis, G. Theocharis, F.K. Diakonos, P. Schmelcher, and Yu.S. Kivshar,
Phys.\ Rev.\ A {\bf 66}, 053608 (2002).

\bibitem{KY}
Yu.S.\ Kivshar and X.\ Yang,
Phys.\ Rev.\ E {\bf 49}, 1657 (1994).

\bibitem{DS-OL1}
%Dark soliton dynamics in spatially inhomogeneous media:
%Application to Bose-Einstein condensates.
G. Theocharis, D.J. Frantzeskakis, R. Carretero-Gonz\'alez,
P.G. Kevrekidis and B.A. Malomed.
Math. Comput. Simulat. {\bf 69} (2005) 537.

\bibitem{DS-OL2}
%Stability of dark solitons in a Bose-Einstein condensate trapped in an optical lattice.
P.G. Kevrekidis, R. Carretero-Gonz\'alez, G. Theocharis,
D.J. Frantzeskakis and B.A. Malomed.
Phys. Rev. A, {\bf 68} 035602 (2003).

\bibitem{DS-OL3}
G.\ Theocharis, D.J.\ Frantzeskakis, R.\ Carretero-Gonz\'alez, P.G.\
  Kevrekidis, and B.A.\ Malomed.
%\newblock {Controlling the motion of dark solitons by means of periodic
%  potentials: Application to Bose-Einstein condensates in optical lattices}.
\newblock {\em Phys.\ Rev.\ E}, {\bf 71} (2005) 017602.

\bibitem{Gross:61}
{E.P.\ Gross}.
%\newblock {Structure of quantized vortex}.
\newblock {\em Nuovo Cim.}, {\bf 20} (1961) 454.

\bibitem{Pitaevskii:61}
{L.P.\ Pitaevskii}.
%\newblock {Vortex lines in an imperfect Bose gas}.
\newblock {\em Sov.\ Phys.\ JETP}, {\bf 13} (1961) 451.

\bibitem{precession1}
A.L.~Fetter,
J.\ Low Temp.\ Phys.\ {\bf 113}, 189 (1998).
\bibitem{precession2}
A.A.~Svidzinsky and A.L.~Fetter,
Phys.\ Rev.\ Lett.\ {\bf 84}, 5919 (2000).
\bibitem{precession3}
E.~Lundh and P.~Ao,
Phys.\ Rev.\ A {\bf 61}, 063612 (2000).
\bibitem{precession4}
J.~Tempere and J.T.~Devreese,
Solid State Comm.\ {\bf 113}, 471 (2000).
\bibitem{precession5}
D.S.~Rokhsar,
Phys.\ Rev.\ Lett.\ {\bf 79}, 2164 (1997).
\bibitem{precession6}
S.A.~McGee and M.J.~Holland,
Phys.\ Rev.\ A {\bf 63}, 043608 (2001).
\bibitem{precession7}
B.P.~Anderson, P.C.~Haljan, C.E.~Wieman and E.A.~Cornell,
Phys.\ Rev.\ Lett.\ {\bf 85}, 2857 (2000).
\bibitem{precession8}
P.O.~Fedichev and G.V.~Shlyapnikov,
Phys.\ Rev.\ A {\bf 60}, R1779 (1999).

\bibitem{Kivshar98}
Y.S.~Kivshar,J.~Christou, V.~Tikhonenko, B.~Luther-Davies, and L.M.~Pismen,
Opt.\ Comm.\ {\bf 152}, 198 (1998).

\bibitem{Parker:04}
%Controlled Vortex-Sound Interactions in Atomic Bose-Einstein Condensates
N.G.~Parker, N.P.~Proukakis, C.F.~Barenghi, and C.S.~Adams,
Phys. Rev. Lett. {\bf 92}, (2004) 160403.

\bibitem{Jackson:98}
B.\ Jackson, J.F.\ McCann, and C.S.\ Adams.
%\newblock {Vortex Formation in Dilute Inhomogeneous Bose-Einstein-Condensates}.
Phys.\ Rev.\ Lett.\ {\bf 80}, 3903 (1998).

\bibitem{Bretin2003a}
% quadrupole oscillation of a single-vortex bose-einstein condensate: evidence for kelvin modes
V.\ Bretin, P.\ Rosenbusch, F.\ Chevy, G.V.\ Shlyapnikov, and J.\ Dalibard,
Phys.\ Rev.\ Lett.\ {\bf 90}, 100403 (2003).

\bibitem{Haljan2001a}
%Use of Surface-Wave Spectroscopy to Characterize Tilt Modes of a Vortex
%in a Bose-Einstein Condensate
P.C.\ Haljan, B.P.\ Anderson, I.\ Coddington, and E.A.\ Cornell,
Phys.\ Rev.\ Lett.\ {\bf 86}, 2922 (2001).

\bibitem{hau} N.S. Ginsberg, J. Brand and L.V. Hau,
Phys. Rev. Lett. {\bf 94}, 040403 (2005).

\end{thebibliography}
\end{document}